\documentclass[aps, prb, superscriptaddress, twocolumn, amssymb, longbibliography]{revtex4-2}

\usepackage{mathtools}
\usepackage{graphicx}

\begin{document}


\title{Optical Conductivity Signatures of Floquet Electronic Phases}

\author{Andrew Cupo}
\email{andrew.cupo@dartmouth.edu}
\affiliation{Department of Physics and Astronomy, Dartmouth College, Hanover, New Hampshire, 03755, USA}

\author{Joshuah T. Heath}
\affiliation{Department of Physics and Astronomy, Dartmouth College, Hanover, New Hampshire, 03755, USA}

\author{Emilio Cobanera}
\affiliation{Department of Physics, SUNY Polytechnic Institute, Utica, New York, 13502, USA}
\affiliation{Department of Physics and Astronomy, Dartmouth College, Hanover, New Hampshire, 03755, USA}

\author{\mbox{James D. Whitfield}}
\affiliation{Department of Physics and Astronomy, Dartmouth College, Hanover, New Hampshire, 03755, USA}

\author{Chandrasekhar Ramanathan}
\affiliation{Department of Physics and Astronomy, Dartmouth College, Hanover, New Hampshire, 03755, USA}

\author{Lorenza Viola}
\email{lorenza.viola@dartmouth.edu}
\affiliation{Department of Physics and Astronomy, Dartmouth College, Hanover, New Hampshire, 03755, USA}


\begin{abstract}
Optical conductivity measurements may provide access to distinct signatures of Floquet electronic phases, which are described theoretically by their quasienergy band structures. We characterize experimental observables of the Floquet graphene antidot lattice (FGAL), which we introduced previously [Phys.\,Rev.\,B \textbf{104}, 174304 (2021)]. On the basis of Floquet linear response theory, the real and imaginary parts of the longitudinal and Hall optical conductivity are computed as a function of probe frequency. We find that the number and positions of peaks in the response function are distinctive of the different Floquet electronic phases, and identify multiple properties with no equilibrium analog. First, for several intervals of probe frequencies, the real part of the conductivity becomes negative. We argue this is indicative of a subversion of the usual Joule heating mechanism: The Floquet drive causes the material to amplify the power of the probe, resulting in gain. Additionally, while the Hall response vanishes at equilibrium, the real and imaginary parts of the Floquet Hall conductivity are non-zero and can be as large as the longitudinal components. Lastly, driving-induced localization tends to reduce the overall magnitude of and to flatten out the optical conductivity signal. From an implementation standpoint, a major advantage of the FGAL is that the above-bandwidth driving limit is reached with photon energies that are at least twenty times lower than that required for the intrinsic material, allowing for significant band renormalization at orders-of-magnitude smaller intensities. Our work provides the necessary tools for experimentalists to map reflectance data to particular Floquet phases for this novel material.
\end{abstract}


\maketitle




\section{Introduction}
\label{sec:intro}

In recent years, Floquet engineering has emerged as a versatile means of producing novel phases of quantum matter with time-periodic external fields \cite{bukov2015universal, basov2017towards, oka2019floquet, de2019floquet, harper2020topology, rudner2020band, rodriguez2021low, Poudel}. For materials subject to external laser irradiation, the polarization, incident photon energy, and intensity can provide independently tunable knobs in the control protocol. Each set of parameters yields a different Floquet electronic phase, which is described theoretically by a corresponding quasienergy band structure. Once these non-equilibrium phases are prepared, their physical properties may be computed by applying Floquet analysis to the time-dependent Schr\"{o}dinger equation \cite{shirley1965solution, sambe1973steady, bukov2015universal, rudner2020floquet}. Based on that, observable quantities should be theoretically determined in order to make contact with experiments. Two techniques that have been utilized are time- and angle-resolved photoemission spectroscopy (TR-ARPES) \cite{wang2013observation, mahmood2016selective, reimann2018subcycle, reutzel2020coherent, keunecke2020electromagnetic, aeschlimann2021survival, zhou2023pseudospin} and DC electronic transport measurements \cite{mciver2020light, candussio2022terahertz}, each with their own benefits and limitations.

In one experiment, the three-dimensional topological insulator Bi$_2$Se$_3$ was subjected to a high intensity and ultrashort 0.124 eV pulse with time-reversal-symmetry-breaking circular polarization \cite{wang2013observation}. The transformation of the equilibrium gapless surface states into gapped Floquet-Bloch bands was detected by TR-ARPES. The precise value of the extracted gap opening of 0.053 eV was found to be consistent with predictions from Floquet theory. A theory paper established that the TR-ARPES bands are located at the quasienergy values, and also determined how the central and side-bands are weighted \cite{fregoso2013driven}. In the TR-ARPES method, the transiently free electrons are also renormalized by the control pulse to form so-called Volkov states \cite{mahmood2016selective}. In a follow-up experiment on the same material, their inclusion was required to explain the ratio of the intensity of the first side-band to the central band as a function of the angle around the $\Gamma$ point \cite{mahmood2016selective}. By now, similar TR-ARPES measurements have been performed on strongly driven Bi$_2$Te$_3$ \cite{reimann2018subcycle}, Cu(111) \cite{reutzel2020coherent}, Au(111) \cite{keunecke2020electromagnetic}, WSe$_2$ \cite{aeschlimann2021survival}, and black phosphorus \cite{zhou2023pseudospin}.

DC electronic transport measurements have also been used to characterize Floquet quantum materials. In particular, graphene was irradiated with a high intensity and ultrashort 0.191 eV pulse with circular polarization \cite{mciver2020light}. Centered at the Dirac point, a 0.060 eV conductance plateau was observed, corresponding to the theoretically predicted gap opening in the Floquet-Bloch bands. Furthermore, a non-zero anomalous Hall conductivity of 1.8 $\pm$ 0.4 $e^2/h$ was detected, which can be explained partially by the generation of topologically non-trivial Floquet-Bloch bands \cite{sato2019microscopic}. A subsequent experimental paper also found a radiation-induced anomalous Hall conductivity in graphene \cite{candussio2022terahertz}.

While TR-ARPES is a powerful technique for detecting dynamically renormalized energy bands, resolving the dispersion over the entire high-symmetry lines remains challenging in practice, due to the fact that the high intensities required restrict pulse lengths to picosecond timescales \cite{wang2013observation, mahmood2016selective, reimann2018subcycle, reutzel2020coherent, keunecke2020electromagnetic, aeschlimann2021survival, zhou2023pseudospin}. Additionally, while DC electronic transport can detect the size of gap openings and topological transitions via jumps in the anomalous Hall conductivity, the dispersion of the quasienergy band structure is not determined \cite{mciver2020light, candussio2022terahertz}. 

\begin{figure*}[t]
\centering{\includegraphics[scale=0.65]{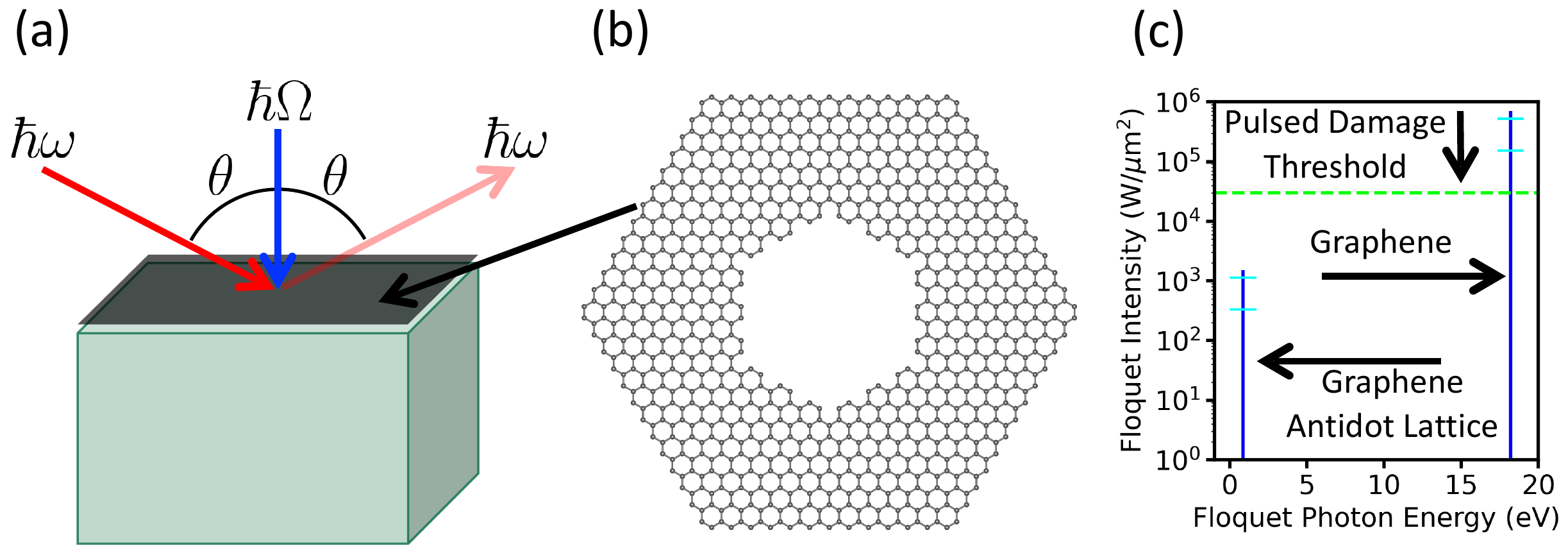}}
\caption{(Color online) {\bf Physical setting under consideration.} {\bf (a)} Schematic of contrast reflectance measurement for few-layer materials \cite{abergel2007visibility, yoo2022spectroscopic}. {\bf (b)} Unit cell of the graphene antidot lattice (GAL), which contains 870 atoms. Notice that the hole is itself another hexagon, where each side length consists of three armchair links. {\bf (c)} The photon energy for the non-resonant above-bandwidth driving limit is more than twenty times greater for pristine graphene as compared to the GAL studied in this work. The blue line on the left represents the range of intensities that are considered for the GAL, all of which are below the pulsed damage threshold \cite{roberts2011response}. The lower and upper cyan dashes correspond to the novel Floquet Dirac and Floquet semi-Dirac phases to be discussed, respectively. The right blue line for pristine graphene encompasses the range of intensities required to obtain the same energy coupling to the Floquet drive as the GAL (mapping also applies between the cyan dashes). We see that there is a large window of intensities that exceed the pulsed damage threshold (horizontal dashed green line), highlighting the significant practical advantages of the Floquet GAL.} 
\label{gal}
\end{figure*}

Optical pump-probe experiments and reflectance measurements have recently emerged as an alternative method of characterizing non-equilibrium phases of materials \cite{earl2021coherent, michael2022generalized, michael2022fresnel}. In the simplest configuration, a Floquet drive with photon energy $\hbar \Omega$ is incident normal to the surface of the material, and an independent probing field with photon energy $\hbar \omega$ reflects off the surface at an angle, see Fig.~\ref{gal}a for a simple schematic. A key benefit of this method is that the same measurement of the ratio of the reflected-to-incident power is simply repeated for different probe frequencies. Theoretically, the relevant quantity to compute in order to make contact with these experiments is the out-of-equilibrium optical conductivity \cite{dehghani2015optical, du2017quadratic, chen2018floquet, kumar2020linear, broers2021observing, eckhardt2022quantum, ahmadabadi2022optical, dabiri2022floquet}. 

Motivated by the above advantages, in this paper we explore the connection between the Floquet quasienergy band structures and the optical conductivity of a novel material we introduced in a previous work, the {\em Floquet graphene antidot lattice} \cite{cupo2021floquet}. We believe there are a number of reasons why this particular system is interesting and well-worth seeking an experimental realization. First, the antidot lattice features geometrically tunable quantum confinement of the electrons while remaining spatially extended in two dimensions. This is to be contrasted with quasi-one-dimensional nanoribbons and nanotubes, and quasi-zero-dimensional quantum dots. Additionally, fabrication of these structures in graphene, at equilibrium, has already been demonstrated using lithographic methods \cite{bai2010graphene, jessen2019lithographic}. Defining atomically precise graphene antidot lattices is now also possible with sophisticated organic-chemistry techniques \cite{moreno2018bottom}. For the Floquet graphene antidot lattice under consideration, the non-resonant above-bandwidth driving limit is reached with near-IR photon energies, which is more than twenty times lower than the same value required for pristine graphene. Under these conditions, obtaining the same energy coupling to the Floquet drive ($C_F$) in both systems requires significantly lower electromagnetic intensities for the graphene antidot lattice. This occurs because $C_F$ is proportional to the ratio of the electric field amplitude to the photon energy (see Eqs. 11 and 12 in \cite{cupo2021floquet}), and because the intensity is proportional to the square of the electric field amplitude. This argument is demonstrated quantitatively in Fig.~\ref{gal}c and constitutes a major practical advantage for pursuing an experimental realization of the Floquet graphene antidot lattice.

The content of the paper is organized as follows. We begin by introducing the tight-binding description of the graphene antidot lattice at equilibrium that will underpin the subsequent analysis (Sec.~\ref{sec:eq}). The external electromagnetic field is then included via the Peierls substitution and the effective electronic properties are determined by applying Floquet analysis to an effective four-band model (Sec.~\ref{sec:floquet}). The necessary mathematical formalism of linear response theory for periodically driven Floquet systems is outlined in Sec.~\ref{sec:linear-response}; in particular, we quote therein a general expression for the optical conductivity in terms of the quasienergies and Floquet eigenstates that will be instrumental for our analysis. Our core numerical results for the longitudinal and Hall optical conductivity are presented and discussed in Sec.~\ref{sec:results-analysis}. We find that the number and positions of the dominant peaks in the optical conductivity vary systematically between the identified Floquet phase regimes and further uncover a number of properties with no equilibrium analog. In Sec.~\ref{sec:results-analysis}, we also include some considerations pertaining to the preparation of Floquet phases from equilibrium, under realistic timing constraints. We conclude with a summary and suggestions for future work in Sec.~\ref{sec:conc}. 

 
\section{Equilibrium Graphene Antidot Lattices}
\label{sec:eq}

The graphene antidot lattice (GAL) is a graphene sheet that has been periodically hole-patterned \cite{pedersen2008graphene}, see Fig.~\ref{gal}b for a representative example. The tight-binding approximation is an established approach for modeling the electronic properties of GALs \cite{pedersen2008graphene, furst2009electronic}. With $i$ and $j$ labeling all sites on the lattice, the second-quantized Hamiltonian takes the simple form
\begin{equation}
H_{\textrm{TB}} = \sum_{ij}^{} t_{ij} a^{\dagger}_i a_j.
\label{tb}
\end{equation}
After performing the lattice Fourier transform, the matrix elements of the Hamiltonian are
\begin{equation}
\big( \mathcal{H}^{(0)}_{\boldsymbol k} \big)_{uv}
= 
\sum_{\boldsymbol R}^{} e^{i \boldsymbol k \cdot \boldsymbol R} t_{u,v\boldsymbol R},
\label{tbdiag}
\end{equation}
where $u$ and $v$ label only the sites within the GAL unit cell, $\boldsymbol R$ is a GAL lattice vector, $t_{u,v\boldsymbol R}$ is the hopping from the site at position $\boldsymbol r_v + \boldsymbol R$ to the site at position $\boldsymbol r_u$, and $\boldsymbol k$ is a wave vector in the GAL first Brillouin zone (BZ). One only needs to consider hopping between nearest neighbors with $t_{u,v\boldsymbol R} = -3.033$ eV. The equilibrium electronic band structure $E_{n \boldsymbol k}$ and the corresponding eigenvectors $\varphi_{n \boldsymbol k}$ are then obtained by solving a standard matrix diagonalization problem,
\begin{equation}
\mathcal{H}^{(0)}_{\boldsymbol k} \varphi_{n \boldsymbol k}
=
E_{n \boldsymbol k} \varphi_{n \boldsymbol k}.
\label{tbsolve}
\end{equation}

It is worth remarking that, whereas we will be working with the above tight-binding model throughout this paper, we used a continuum Dirac Hamiltonian approach in our previous study \cite{cupo2021floquet}. Both models have been employed in the literature, and a comparison is reported in \cite{brun2014electronic}. In hindsight, while the Dirac method is seemingly simpler, in practice it can demand significantly more computational effort. This largely stems from the fact that the wave functions were expanded in two-dimensional (2D) spatial Fourier series with many components, yielding large Hamiltonian matrices with mostly non-zero elements. In contrast, the number of rows in the matrix representation of the tight-binding Hamiltonian is simply equal to the number of atoms in the unit cell (870 for the case at hand). Also, the matrix elements are easy to compute from Eq.~\ref{tbdiag}, with most being zero. In any case, the Dirac Hamiltonian for pristine graphene is a low-energy approximation obtained by Taylor-expanding the tight-binding solution about the K point. Therefore, the tight-binding technique contains more physics by design and is thus expected to be a better reflection of Nature. Lastly, we do not include spin-orbit coupling effects since the band energy corrections would only be on the order of tens of $\mu$eV \cite{gmitra2009band, konschuh2010tight}.


\section{Floquet Graphene Antidot Lattices}
\label{sec:floquet}

Within a tight-binding formalism, periodic driving by an external electromagnetic field may be included by making the so-called Peierls substitution \cite{li2020electromagnetic}. The result is the \textit{full} time-dependent Hamiltonian
\begin{equation}
\mathcal{H}_{\boldsymbol k}(t) 
= 
\mathcal{H}^{(0)}_{\boldsymbol k}(t_{u,v\boldsymbol R} \, \mapsto \, t_{u,v\boldsymbol R} \, e^{i \theta_{u,v\boldsymbol R}(t)}),
\label{peierls}
\end{equation} 
where
\begin{equation}
\theta_{u,v\boldsymbol R}(t)
=
\frac{-|e|}{\hbar} \int_{\boldsymbol r_v + \boldsymbol R}^{\boldsymbol r_u} d \boldsymbol r \cdot \boldsymbol A(t).
\label{angle}
\end{equation} 
We consider a vector potential with time-reversal-symmetry-breaking circular polarization,
\begin{equation}
\boldsymbol A(t) = \frac{E_0}{\Omega} [\textrm{cos}(\Omega t),\textrm{sin}(\Omega t),0],
\label{vectorpot}
\end{equation} 
with $E_0$ and $\Omega$ being the electric field amplitude and the angular frequency, respectively, and we assume that the GAL is located in the $z=0$ plane (see Fig.\,\ref{gal}b). 

At this point it is convenient to define two additional operators, specifically, the \textit{full} linear electronic current operator, namely,
\begin{equation}
\mathcal{J}_{\boldsymbol k}^{(\alpha)}(t) 
\equiv 
\frac{-|e|}{\hbar}
\frac{\partial \mathcal{H}_{\boldsymbol k}(t)}{\partial k_\alpha},
\label{fullcurrent}
\end{equation}
and the \textit{full} inverse effective mass tensor operator,
\begin{equation}
\mathcal{M}_{\boldsymbol k}^{(\alpha \beta)}(t) 
\equiv 
\frac{1}{\hbar^2}
\frac{\partial^2 \mathcal{H}_{\boldsymbol k}(t)}{\partial k_\alpha \partial k_\beta},
\label{fullmass}
\end{equation}
with $\alpha, \beta \in \{x,y,z\}$. Analytical expressions for these two operators can be obtained based on Eqs.~\ref{tbdiag} and~\ref{peierls}-\ref{vectorpot}.

While solving the equilibrium problem produces the same number of bands as there are atoms in the unit cell (870, as noted before), the essential physics is captured by the bands closest to the Fermi energy (0 eV in plots to be discussed). In particular, for reasons that will become clearer later on, one should retain in the analysis the first two valence bands and the first two conduction bands for our system. The matrix elements of the 4$\times$4 \textit{reduced} time-dependent Hamiltonian are then obtained as
\begin{equation}
\big(H_{\boldsymbol k}(t)\big)_{nn'}
=
\varphi_{n \boldsymbol k}^\dagger 
\mathcal{H}_{\boldsymbol k}(t) 
\varphi_{n' \boldsymbol k}.
\label{hamred}
\end{equation}
In similar fashion, we obtain the matrix elements of the \textit{reduced} linear electronic current operator
\begin{equation}
\big(j_{\boldsymbol k}^{(\alpha)}(t)\big)_{nn'}
=
\varphi_{n \boldsymbol k}^\dagger 
\mathcal{J}_{\boldsymbol k}^{(\alpha)}(t) 
\varphi_{n' \boldsymbol k}
\label{currentred}
\end{equation}
and the \textit{reduced} inverse effective mass tensor operator
\begin{equation}
\big(\mu_{\boldsymbol k}^{(\alpha \beta)}(t)\big)_{nn'}
=
\varphi_{n \boldsymbol k}^\dagger 
\mathcal{M}_{\boldsymbol k}^{(\alpha \beta)}(t)
\varphi_{n' \boldsymbol k}.
\label{massred}
\end{equation}
It is worth remarking that one might instead attempt to obtain $j_{\boldsymbol k}^{(\alpha)}(t)$ and $\mu_{\boldsymbol k}^{(\alpha \beta)}(t)$ by taking the $k$-derivatives directly. However, notice that this would implicitly require taking numerical $k$-derivatives of wave functions via Eq.~\ref{hamred}, which is not possible using standard finite differences since each eigenvector can be multiplied by an arbitrary complex number of magnitude one. Thus, Eqs.~\ref{currentred} and~\ref{massred} must be used to consistently define $j_{\boldsymbol k}^{(\alpha)}(t)$ and $\mu_{\boldsymbol k}^{(\alpha \beta)}(t)$. The dimensional restriction from 870 to four bands makes the problem at hand computationally tractable, whereas the equilibrium Hamiltonians in other Floquet studies already contained few-bands and no such procedure was required \cite{dehghani2015optical, du2017quadratic, chen2018floquet, kumar2020linear, broers2021observing, eckhardt2022quantum, ahmadabadi2022optical, dabiri2022floquet}.

Following an initialization step during which the applied electric field is turned on (see Sec.\,\ref{sec:results-analysis}D), the Hamiltonian depends periodically on time, $H_{\boldsymbol k}(t+T) = H_{\boldsymbol k}(t)$, with $T = 2 \pi/\Omega$. We can thus construct a complete set of solutions to the time-dependent Schr\"{o}dinger equation,  
\begin{equation}
i \hbar \partial_t \psi_{n \boldsymbol k}(t) = H_{\boldsymbol k}(t) \psi_{n \boldsymbol k}(t),
\label{tdse}
\end{equation} 
by using Floquet's factorization Ansatz, namely, 
\begin{equation}
\psi_{n \boldsymbol k}(t) = e^{-i \epsilon_{n \boldsymbol k} t/\hbar} \Phi_{n \boldsymbol k}(t), \quad
\Phi_{n \boldsymbol k}(t+T) = \Phi_{n \boldsymbol k}(t),
\label{floquetsol}
\end{equation}
where $\epsilon_{n \boldsymbol k} \in {\mathbb R}$ are the {\em Floquet quasienergies}. Due to the time-periodicity of the Hamiltonian, there exists a redundancy in the solutions, since any quasienergy may be shifted by arbitrary integer multiples of $\hbar \Omega$ without the corresponding Floquet state changing, 
\begin{eqnarray}
\psi_{n \boldsymbol k}(t) & = &e^{-i \epsilon_{n \boldsymbol k} t/\hbar} \Phi_{n \boldsymbol k}(t)  \nonumber 
\\
& = &e^{-i (\epsilon_{n \boldsymbol k} + \ell \hbar \Omega) t/\hbar} (e^{i \ell \Omega t} \Phi_{n \boldsymbol k}(t)) \nonumber
\\
& \equiv &e^{-i \epsilon_{n \boldsymbol k, \ell} t/\hbar} \Phi_{n \boldsymbol k, \ell}(t) 
\\
& \equiv &\psi_{n \boldsymbol k, \ell}(t), \quad \ell \in {\mathbb Z}. 
\label{redundant}
\end{eqnarray}
However, in analogy with the emergence of a crystal momentum BZ for a particle in a spatially periodic potential, all {\em distinct} Floquet state solutions can be labeled by quasienergies that fall within a single {\em Floquet BZ} of width $\hbar \Omega$. In particular, the {\em first} Floquet BZ is defined by requiring that 
\begin{equation}
-\hbar \Omega/2 \leq \epsilon_{n \boldsymbol k, \tilde{\ell}(n, \boldsymbol k)} < \hbar \Omega / 2.
\label{firstFBZ}
\end{equation}
Quasienergies restricted to a single Floquet BZ are often  referred to as {\em folded} in the literature, as opposed to {\em unfolded} ones, whose values are unrestricted. Depending on the driving parameters, care is needed in establishing a correspondence between folded and unfolded quasienergies, due to the possibility that different copies of the equilibrium band structure overlap \cite{rudner2020floquet, vogl2020unfold}. While a fairly general and useful approach to unfold the quasienergy spectrum is to compute the time-averaged spectral function \cite{rudner2020floquet},  this is unnecessary in the regime of above-bandwidth driving we are interested in here. Specifically, by assuming that the driving energy is larger than the range of the equilibrium energy bands,
\begin{equation}
\hbar \Omega >  \textrm{max}\{E_{n \boldsymbol k}\} - \textrm{min}\{E_{n \boldsymbol k}\}, \quad \forall n, {\boldsymbol k},
\label{extent}
\end{equation}
we may consider the quasienergy bands to be unfolded when the equilibrium energies are recovered in the limit of vanishing driving amplitude: 
\begin{equation}
\lim_{E_0 \to 0} \epsilon_{n \boldsymbol k, 0}
\equiv
\lim_{E_0 \to 0} \epsilon_{n \boldsymbol k} 
= 
E_{n \boldsymbol k}.
\label{unfold}
\end{equation}
In this way, the folded quasienergies that are automatically obtained from numerical diagonalization and the unfolded quasienergies are guaranteed to coincide.

Because $H_{\boldsymbol k}(t)$ and $\Phi_{n \boldsymbol k}(t)$ are time-periodic, we have the following Fourier decompositions
\begin{equation}
H_{\boldsymbol k}(t) = \sum_{m}^{} H_{\boldsymbol k}^{(m)} e^{-i m \Omega t},
\label{hamfourier}
\end{equation}
\begin{equation}
\Phi_{n \boldsymbol k}(t) = \sum_{m}^{} \phi_{n \boldsymbol k}^{(m)} e^{-i m \Omega t},
\label{phifourier}
\end{equation}
where $m\in {\mathbb Z}$. Eq.~\ref{tdse} is thus transformed into
\begin{equation}
\sum_{m'}^{} 
\tilde H_{\boldsymbol k}^{(m,m')}
\phi_{n \boldsymbol k}^{(m')} 
= 
\epsilon_{n \boldsymbol k} 
\phi_{n \boldsymbol k}^{(m)}, 
\label{floquetsolve}
\end{equation}
in terms of 
\begin{eqnarray}
\tilde H_{\boldsymbol k}^{(m,m')}
& = &
\frac{1}{T} \int_{0}^{T} dt H_{\boldsymbol k}(t) e^{i (m-m') \Omega t}
-
\delta_{mm'} m \hbar \Omega \boldsymbol 1
\nonumber \\
& = &
H_{\boldsymbol k}^{(m-m')}
-
\delta_{mm'} m \hbar \Omega \boldsymbol 1, 
\label{floquetdiag}
\end{eqnarray}
where $\boldsymbol 1$ denotes the identity matrix. In this way, the original time-dependent problem has been converted into a time-independent problem of diagonalizing a Hermitian operator in an extended, in principle infinite-dimensional, Hilbert space \cite{sambe1973steady, bukov2015universal, rudner2020floquet}. In practice, for all cases to be considered, the quasienergy spectra $\epsilon_{n \boldsymbol k}$ are converged with $N_{\textrm{Floquet}} = 7$ temporal harmonics. Accordingly, the sum over $m'$ in Eq.~\ref{floquetsolve} ranges from $-N_{\textrm{Floquet}}$ to $N_{\textrm{Floquet}}$. Within this formalism, we proceed to model the Floquet graphene antidot lattice (FGAL).


\section{Floquet Linear Response Theory: Optical Conductivity}
\label{sec:linear-response}

Equilibrium linear response theory has been extended and applied to Floquet quantum materials in multiple studies \cite{torres2005kubo, oka2009photovoltaic, dehghani2015out, dehghani2015optical, oka2016heterodyne,chen2018floquet, wackerl2020floquet, kumar2020linear, rudner2020floquet, ahmadabadi2022optical}. In essence, the formalism determines how the expectation value of the total electronic current changes when a weak probe field is added to the Floquet system, say,
\begin{equation}
\delta \left\langle \boldsymbol J \right\rangle(t)
\equiv
\textrm{Tr}
\{
\rho'(t) \boldsymbol J'(t)
-
\rho(t) \boldsymbol J(t)
\}.
\label{expcur}
\end{equation}
Here, $\rho$ ($\rho'$) and $\boldsymbol J$ ($\boldsymbol J'$) are the density matrix and the total electronic current operator, respectively, for the Floquet system without (including) the weak probing electric field at angular frequency $\omega$,
\begin{equation}
\boldsymbol E_{\textrm{pro}}(t) 
=
e^{-i \omega t} 
\boldsymbol E_{0,\textrm{pro}}.
\label{probe}
\end{equation}
The probe is assumed to act on the established Floquet system starting at time $t_0$, which can be at any time during a driving cycle. The expectation value in Eq.~\ref{expcur} is then determined at some later time $t > t_0$, with the calculation being performed in the interaction picture with respect to the unperturbed Floquet Hamiltonian:
\begin{equation}
\delta \left\langle \boldsymbol J \right\rangle(t)
= 
\textrm{Tr}
\{
\rho'^{(I)}(t) \boldsymbol J'^{(I)}(t)
-
\rho^{(I)}(t) \boldsymbol J^{(I)}(t)
\}.
\label{expcurint}
\end{equation}
Note that operators may be expanded in the single-particle basis as follows
\begin{equation}
\mathcal{A}(t)
=
\sum_{\boldsymbol{k}nn'}^{} 
c^{\dagger}_{n \boldsymbol k}
[\mathcal{A}_{\boldsymbol k}(t)]_{nn'}
c_{n' \boldsymbol k},
\label{singpart}
\end{equation}
where $\mathcal{A}(t)$ is an arbitrary operator, and the $c^{\dagger}_{n \boldsymbol k}$ ($c_{n' \boldsymbol k}$) are creation (annihilation) operators for the Floquet eigenstates. The transformation to the interaction picture is then effected by
\begin{equation}
\mathcal{A}^{(I)}(t)
\equiv
U^{\dagger}(t,t_0)
\mathcal{A}(t)
U(t,t_0),
\label{inttran}
\end{equation}
\begin{equation}
\mathcal{A}^{(I)}_{\boldsymbol k}(t)
\equiv
U^{\dagger}_{\boldsymbol k}(t,t_0)
\mathcal{A}_{\boldsymbol k}(t)
U_{\boldsymbol k}(t,t_0),
\label{inttran_k}
\end{equation}
where the time evolution operators are given by 
\begin{equation}
U(t,t_0)
=
\mathcal{T}
\textrm{exp}
\bigg\{
-\frac{i}{\hbar}
\int_{t_0}^{t} ds H(s) 
\bigg\},
\label{quantimeevol}
\end{equation}
\begin{equation}
U_{\boldsymbol k}(t,t_0)
=
\mathcal{T}
\textrm{exp}
\bigg\{
-\frac{i}{\hbar}
\int_{t_0}^{t} ds H_{\boldsymbol k}(s) 
\bigg\},
\label{quantimeevol_k}
\end{equation}
with $\mathcal{T}$ denoting time-ordering.    

The \textit{linear} in linear response comes from the fact that the time evolution of the density matrix in the interaction picture,
\begin{equation}
i \hbar \partial_t \rho'^{(I)}(t) 
=
[H^{(I)}_{\textrm{pro}}(t),\rho'^{(I)}(t)],
\label{denmat}
\end{equation}
is approximated to the lowest, linear order as
\begin{equation}
\rho'^{(I)}(t) 
\approx
\rho(t_0)
-
\frac{i}{\hbar}
\int_{t_0}^{t} ds
[H^{(I)}_{\textrm{pro}}(s),\rho(t_0)],
\label{lindenmat}
\end{equation}
where $H^{(I)}_{\textrm{pro}}(t)$ is the component of the Hamiltonian containing the probe, in the interaction picture. At this point, Eqs.~\ref{singpart}, \ref{inttran_k}, and \ref{lindenmat} are used to expand Eq.~\ref{expcurint}. In the process, 
\begin{equation}
f_{n \boldsymbol k}
\equiv
\textrm{Tr}
\{
\rho(t_0)
c^{\dagger}_{n \boldsymbol k}
c_{n \boldsymbol k}
\}
\vspace*{2mm}
\label{occdef}
\end{equation}
emerges as the expectation value of the (fermionic) occupation number for the Floquet eigenstates with respect to the state at time $t_0$, which we assume to be diagonal in the Floquet basis. For \textit{Floquet} linear response theory, an essential step is to leverage the exact expansion of the propagator defined by Eq.~\ref{quantimeevol_k} in terms of the Floquet unfolded quasienergies and eigenstates,
\begin{equation}
U_{\boldsymbol k}(t,t_0)
=
\sum_{n}^{}
e^{-i \epsilon_{n \boldsymbol k} (t-t_0) / \hbar}
\Phi_{n \boldsymbol k}(t)
\Phi^{\dagger}_{n \boldsymbol k}(t_0). 
\label{expandprop}
\end{equation}
Ultimately, Ohm's law for the induced current emerges,
\begin{equation}
\delta \left\langle \boldsymbol J \right\rangle(t)
= 
\sigma(t) 
\boldsymbol E_{\textrm{pro}}(t),
\label{ohm}
\end{equation}
with the conductivity tensor
\begin{equation}
\sigma(t) = \sum_{N}^{} \sigma_N(\omega) e^{-i N \Omega t}, \quad N \in {\mathbb Z}.
\label{condfourier}
\end{equation}

The above highlights that the conductivity of a Floquet system is {\em itself} now time-dependent, with the same periodicity as the drive. In our analysis, we shall focus on the homodyne ($N = 0$) component since the heterodyne ($N \neq 0$) contributions are much smaller when the Floquet photon energy is above the total equilibrium bandwidth \cite{kumar2020linear}. The explicit expression for the Floquet optical conductivity is \cite{kumar2020linear}
\begin{widetext}
\begin{align}
&\sigma_N^{(\alpha \beta)}(\omega)\notag\\
&=\frac{i}{\omega}
\int_{\textrm{BZ}}^{} \frac{d^\mathcal{D} k}{(2 \pi)^{\mathcal{D}}}
\sum_{n}^{} 
f_{n \boldsymbol k}
\Bigg[
\sum_{n'm}^{}
\Bigg(
\frac{\mathcal{F}_{(n,-N;n',m)\boldsymbol k}[j_{\boldsymbol k}^{(\alpha)}(t)] \mathcal{F}_{(n',m;n,0)\boldsymbol k}[j_{\boldsymbol k}^{(\beta)}(t)]}{\hbar \omega + ( \epsilon_{n \boldsymbol k} - \epsilon_{n' \boldsymbol k} - m \hbar \Omega ) + i \eta}
-
\frac{\mathcal{F}_{(n,0;n',m)\boldsymbol k}[j_{\boldsymbol k}^{(\beta)}(t)] \mathcal{F}_{(n',m;n,N)\boldsymbol k}[j_{\boldsymbol k}^{(\alpha)}(t)]}{\hbar \omega - ( \epsilon_{n \boldsymbol k} - \epsilon_{n' \boldsymbol k} - m \hbar \Omega ) + i \eta}
\Bigg)\notag\\
& \phantom{=}+
e^2 \mathcal{F}_{(n,0;n,N)\boldsymbol k}[\mu_{\boldsymbol k}^{(\alpha \beta)}(t)]
\Bigg], \qquad \alpha,\beta \in \{x,y,z\}, \quad \eta >0,  
\label{cond}
\end{align}
\end{widetext}
where $\mathcal{D}$ denotes the spatial dimension (here, ${\mathcal D}=2$) and the generalized Floquet matrix elements of a time-dependent operator $A_{\boldsymbol k}(t)$ are understood as 
\begin{equation}
\begin{split}
& \mathcal{F}_{(n,m;n',m')\boldsymbol k}[A_{\boldsymbol k}(t)] \\
& \equiv \frac{1}{T} \int_{0}^{T} dt e^{i (m'-m) \Omega t} \Phi^\dagger_{n \boldsymbol k}(t) A_{\boldsymbol k}(t) \Phi_{n' \boldsymbol k}(t).
\end{split}
\label{floquetmatel}
\end{equation}
Since, in arriving at Eq.\,\ref{cond}, only the diagonal elements of the density matrix in the Floquet basis are assumed to contribute to the conductivity and the initial state is diagonal, any dependence upon $t_0$ drops out. It is also worth noting that, formally, the regularization parameter $\eta$ should be taken to zero at the end of the treatment; however, in practice, it phenomenologically accounts for the effective electron quasiparticle linewidth (inverse lifetime) due to electron-electron and electron-phonon interactions, defects, and impurities \cite{torres2005kubo}. As will be apparent from our numerical simulations, $\eta$ broadens the peaks in the optical conductivity, creating a non-local ``tail effect'' that has a different behavior in the real and imaginary parts. The numerical evaluation of Eq.~\ref{cond} is highly non-trivial; we break down the required steps in Appendix~\ref{sec:comp}, while highlighting the associated computational costs.

We point out that $\boldsymbol E_{\textrm{pro}}(t)$ and $\delta \left\langle \boldsymbol J \right\rangle(t)$ are in complex form for mathematical convenience only, and that the physical quantities are obtained by taking the real part. In particular, we have
\begin{eqnarray}
\label{realprobe}
\boldsymbol E_{\textrm{pro,physical}}(t) 
& \equiv & 
\textrm{Re}\{ \boldsymbol E_{\textrm{pro}}(t) \} 
= 
\textrm{cos}(\omega t) \boldsymbol E_{0,\textrm{pro}} 
\\
\label{realcurrent}
\delta \left\langle \boldsymbol J \right\rangle_\textrm{physical}(t)
& \equiv &
\textrm{Re}\{ \delta \left\langle \boldsymbol J \right\rangle(t) \} 
\\ 
& = & 
\sum_{N}^{} \Big[ \textrm{Re}\{\sigma_N(\omega)\} \textrm{cos}((\omega + N \Omega) t) \nonumber
\\
& + & \textrm{Im}\{\sigma_N(\omega)\} \textrm{sin}((\omega + N \Omega) t) \Big] \boldsymbol E_{0,\textrm{pro}}. \nonumber
\end{eqnarray}

As noted, in this paper we focus on the optical conductivity as a function of {\em non-zero} probe frequencies that connects to reflectance measurements. For completeness, for the reader interested in DC electronic transport, we include in Appendix~\ref{sec:dc} a summary of the relevant zero-frequency conductivity formulas.


\section{Results and Analysis}
\label{sec:results-analysis}

\begin{figure}[t]
\centering{\includegraphics[scale=1]{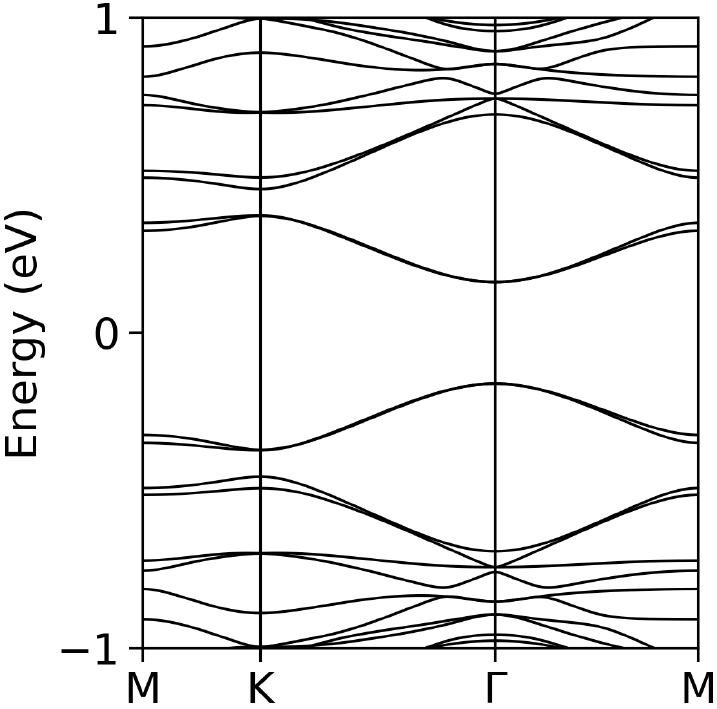}}
\caption{{\bf Equilibrium electronic band structure of the GAL shown in Fig.~\ref{gal}b.} Tight-binding calculations based on Eq.~\ref{tbsolve} are carried out with the Fermi energy at 0 eV. Whereas pristine graphene features the Dirac dispersion, quantum confinement opens an electronic band gap of 0.32 eV and leads to quadratic bands at $\Gamma$.}
\label{bands}
\end{figure}

\begin{figure*}[t]
\includegraphics[scale=0.9]{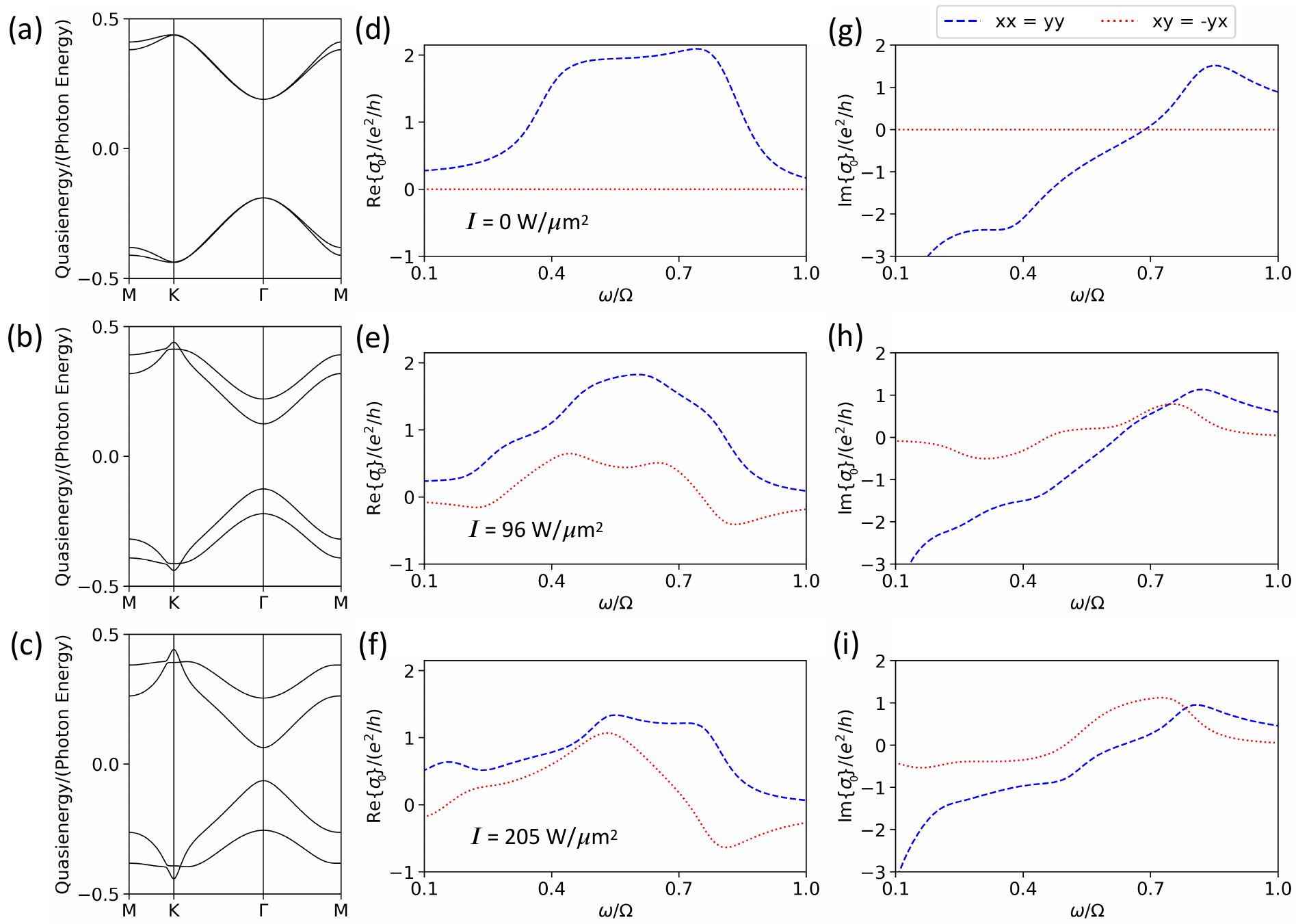}
\caption{(Color online) {\bf Quasienergy band structures and optical conductivity sweeps in the Floquet quasi-equilibrium phase regime.} $E_0$ is the electric field amplitude in units where 1 a.u. = 1.291 (GeV/nm)/nC \cite{hubener2017creating,cupo2021floquet} and $I$ is the corresponding intensity $I = (\epsilon_0 c/2) E_0^2$. {\bf (a)} $E_0 =$ 0.00 a.u. and $I =$ 0 W/$\mu$m$^2$. Equilibrium limit, which maps directly to Fig.~\ref{bands}. {\bf (b)} $E_0 =$ 1.30 a.u. and $I =$ 96 W/$\mu$m$^2$. {\bf (c)} $E_0 =$ 1.90 a.u. and $I =$ 205 W/$\mu$m$^2$. {\bf (d-f)} Corresponding real parts of the longitudinal and Hall homodyne conductivity. $\omega/\Omega$ is the ratio of the probe to driving angular frequency. {\bf (g-i)} Corresponding imaginary parts of the longitudinal and Hall homodyne conductivity. Note that, in practice, the relatively large intensities of the driving field are only sustained for picosecond timescales.}
\label{oc1}
\end{figure*}

\begin{figure*}[t]
\includegraphics[scale=0.9]{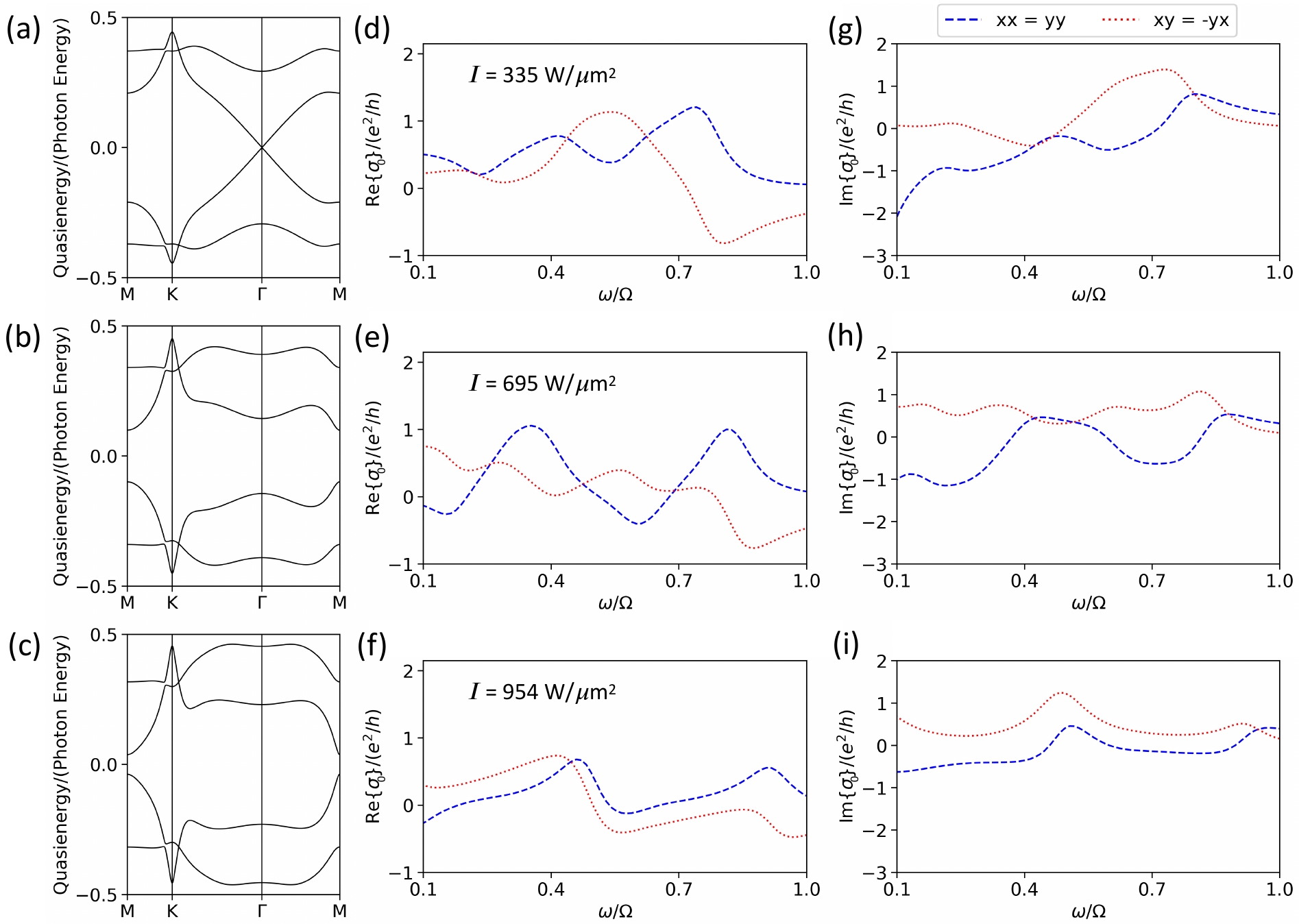}
\caption{(Color online) {\bf Quasienergy band structures and optical conductivity sweeps in the Floquet Dirac phase regime.} $E_0$ is the electric field amplitude in units where 1 a.u. = 1.291 (GeV/nm)/nC \cite{hubener2017creating,cupo2021floquet} and $I$ is the corresponding intensity $I = (\epsilon_0 c/2) E_0^2$. {\bf (a)} $E_0 =$ 2.43 a.u. and $I =$ 335 W/$\mu$m$^2$. Dynamical restoration of the Dirac dispersion. {\bf (b)} $E_0 =$ 3.50 a.u. and $I =$ 695 W/$\mu$m$^2$. The quasienergy gap is shifted to M. {\bf (c)} $E_0 =$ 4.10 a.u. and $I =$ 954 W/$\mu$m$^2$. Onset of selective dynamical localization. {\bf (d-f)} Corresponding real parts of the longitudinal and Hall homodyne conductivity. $\omega/\Omega$ is the ratio of the probe to driving angular frequency. {\bf (g-i)} Corresponding imaginary parts of the longitudinal and Hall homodyne conductivity.}
\label{oc2}
\end{figure*}

\begin{figure*}[t]
\includegraphics[scale=0.9]{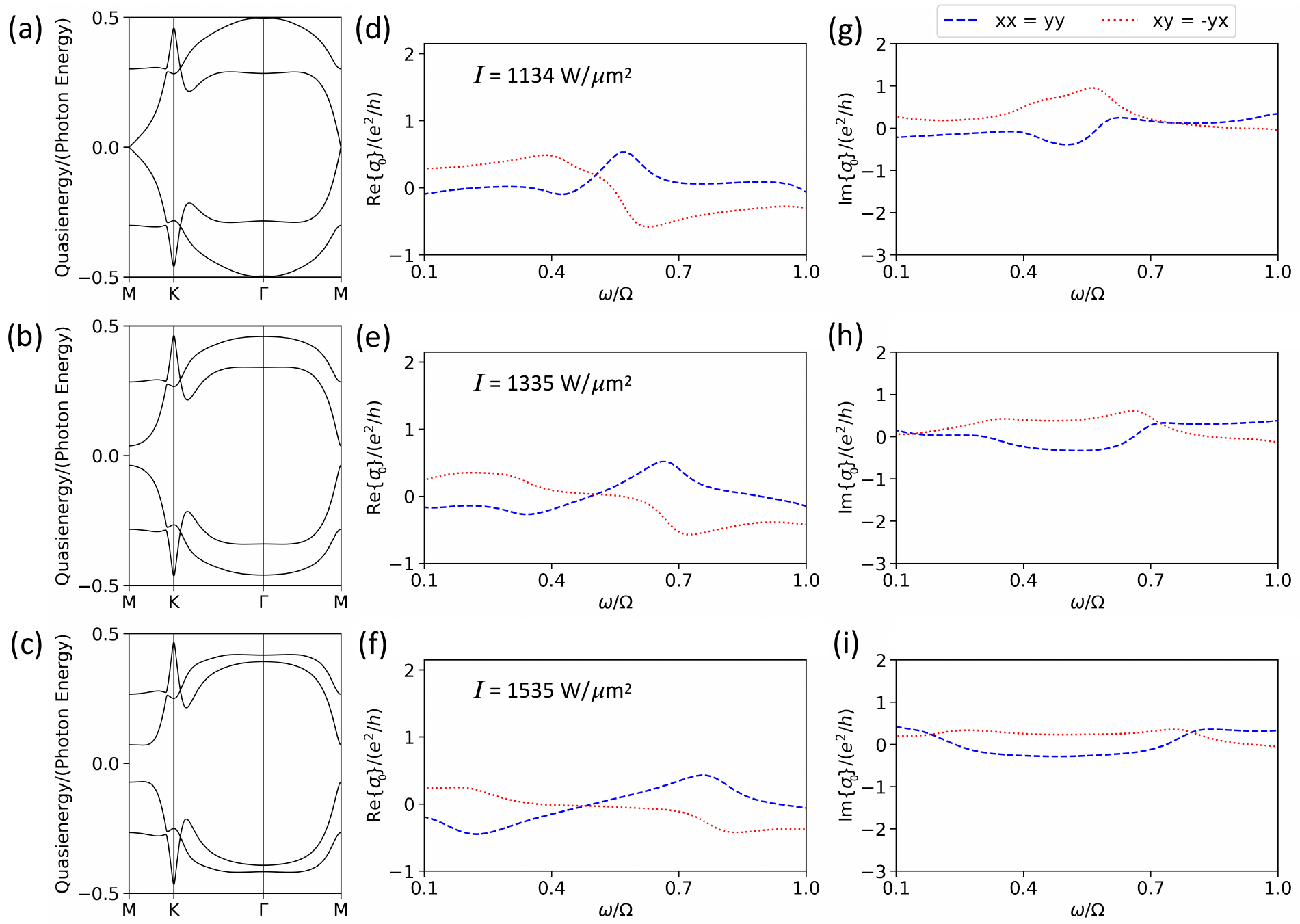}
\caption{(Color online) {\bf Quasienergy band structures and optical conductivity sweeps in the Floquet semi-Dirac phase regime.} $E_0$ is the electric field amplitude in units where 1 a.u. = 1.291 (GeV/nm)/nC \cite{hubener2017creating,cupo2021floquet} and $I$ is the corresponding intensity $I = (\epsilon_0 c/2) E_0^2$. {\bf (a)} $E_0 =$ 4.47 a.u. and $I =$ 1134 W/$\mu$m$^2$. Emergence of the Floquet semi-Dirac phase. From M to K the inner bands are quadratic and from M to $\Gamma$ the inner bands are linear. {\bf (b)} $E_0 =$ 4.85 a.u. and $I =$ 1335 W/$\mu$m$^2$. (c) $E_0 =$ 5.20 a.u. and $I =$ 1535 W/$\mu$m$^2$. {\bf (d-f)} Corresponding real parts of the longitudinal and Hall homodyne conductivity. $\omega/\Omega$ is the ratio of the probe to driving angular frequency. {\bf (g-i)} Corresponding imaginary parts of the longitudinal and Hall homodyne conductivity.}
\label{oc3}
\end{figure*}

Taking into consideration both the geometry of the GAL and the choice of the driving protocol, there is a priori a large parameter space to consider. In practice, we only vary the driving (electric field) amplitude and fix the other variables to values that we expect to be accessible experimentally. First, the structure is chosen to be the one shown in Fig.~\ref{gal}b, which is a triangular GAL with 870 atoms per unit cell. The even number of atoms in the unit cell and the edge morphology ensure that there are no edge states to consider \cite{brun2014electronic}. Also, note that this structure matches closely with one of the systems we studied previously ($L$ = 3.0 nm and $D$ = 1.6 nm in \cite{cupo2021floquet}).

The equilibrium electronic band structure defined by Eq.~\ref{tbsolve} is shown in Fig.~\ref{bands} for this system. The introduction of the antidot lattice opens a large gap of 0.32 eV at the superlattice $\Gamma$ point. In order to make the Floquet calculations practical, as mentioned in Sec.\,\ref{sec:floquet}, restriction to only the most important bands near the Fermi energy (0 eV) is necessary. We see from Fig.~\ref{bands} that the first two valence bands are nearly degenerate along large sections of the high-symmetry lines, so that both should be considered. Due to particle-hole symmetry, the same is true of the first two conduction bands. Therefore, as anticipated, we consider a reduced four-band system. We may further understand intuitively why this is the case. In our previous study \cite{cupo2021floquet}, the Dirac model only contained the K point of the intrinsic material, resulting in two total bands folded onto the $\Gamma$ point of the superlattice. However, the tight-binding framework automatically contains the K and K$'$ points of pristine graphene, leading to four total bands folded onto the $\Gamma$ point of the antidot lattice.

The driving protocol is based on the vector potential given in Eq.~\ref{vectorpot}. We consider circular polarization and $\hbar \Omega =$ 0.85 eV. The photon energy is chosen to be small enough so that transitions into neglected bands are not possible, yet large enough as to include the four equilibrium bands under consideration. This ensures that the quasienergy bands in the first Floquet BZ (between $-\hbar \Omega/2$ and $\hbar \Omega/2$) are unfolded, as in Eq.~\ref{unfold}. The effect of tuning the Floquet photon energy may be considered in future work. 

In the numerics, the electric field amplitude $E_0$ was varied from 0.00 a.u. to 10.00 a.u. in increments of 0.05 a.u., with interpolations as necessary, and the quasienergy band structure was computed for each case. The relevant conversion here is 1 a.u. = 1.291 (GeV/nm)/nC, as originally defined in \cite{hubener2017creating} and also utilized in our previous paper \cite{cupo2021floquet}. In the figure captions to follow, we further report the intensity $I = (\epsilon_0 c/2) E_0^2$ in physical units that would be useful to an experimentalist. 

We identified nine particularly interesting Floquet electronic phases and computed the corresponding optical conductivity sweep for each. We estimate $\eta =$ 0.05 eV based on previous first-principles calculations of the electron linewidth for graphene due to electron-electron and electron-phonon scattering \cite{park2009first}. We consider ideal occupation of the Floquet eigenstates ($f_{1 \boldsymbol k} = f_{2 \boldsymbol k} = 1$ and $f_{3 \boldsymbol k} = f_{4 \boldsymbol k} = 0$) \cite{dehghani2015out, dehghani2015optical, kumar2020linear, d2015dynamical}. Lastly, we find that transforming the hexagonal BZ into a rectangle and using a $101 \times 101$ k-grid ensures converged BZ integrals.


\subsection{Quasienergy Band Structures of GAL Floquet Phases}

The quasienergy band structures are plotted in the left column (panels a, b, and c) of Figs.~\ref{oc1}-\ref{oc3}. Fig.~\ref{oc1}a contains the equilibrium limit ($E_0 =$ 0 a.u.) of the quasienergy bands. We notice that there is a direct map to the electronic band structure in Fig.~\ref{bands}, which indicates that the plotted quasienergy bands are unfolded (Eq.~\ref{unfold}). In general, we notice across all plots that the driving tends to split the bands that were nearly degenerate at equilibrium. As the amplitude of the driving increases in going to Figs.~\ref{oc1}b and~\ref{oc1}c, the energy gap at $\Gamma$ decreases with continuous tunability.

In Fig.~\ref{oc2}a the energy gap has closed completely, restoring the Dirac dispersion. As $E_0$ increases further, the energy gap at $\Gamma$ reopens and shifts to the M point in Fig.~\ref{oc2}b. This behavior continues in Fig.~\ref{oc2}c, except that we also observe the onset of {\em selective dynamical localization} (SDL) \cite{cupo2021floquet}, i.e., the bands near $\Gamma$ flatten. The SDL persists in the remaining quasienergy spectra in Fig.~\ref{oc3}.  

We further observe the emergence of the novel {\em Floquet semi-Dirac phase} in Fig.~\ref{oc3}a, where the gap closes at M and the bands emanating from this point feature quadratic (M-K) and linear (M-$\Gamma$) dispersions \cite{delplace2013merging, agarwala2016effects, liu2019engineering, cupo2021floquet}. This indicates the co-existence of non-relativistic-like and relativistic-like quasiparticles, respectively. In the final two panels, Figs.~\ref{oc3}b and~\ref{oc3}c, the energy gap reopens at M and the flattened regions near $\Gamma$ shift outwards. Interestingly, in the final band structure (Fig.~\ref{oc3}c), the inner and outer bands approach each other again as they had at equilibrium, except now the bands are all significantly flattened near $\Gamma$.

The quasienergy band structures were intentionally organized into three groups of three in the above description. The three figures correspond to different Floquet electronic phase regimes as follows: (i) Floquet quasi-equilibrium (Fig.~\ref{oc1}); (ii) Floquet Dirac (Fig.~\ref{oc2}); and (iii) Floquet semi-Dirac (Fig.~\ref{oc3}). Accordingly, the first phase regime starts at equilibrium by definition, whereas the other two begin with the Floquet electronic phases where the driving has closed the quasienergy gap (Floquet Dirac and Floquet semi-Dirac).

On a quick inspection, it appears as though the two valence bands approach each other at two momentum values near the K point in general. By plotting the quasienergy bands on a fine grid, we find that there is an avoided crossing near K along the K-M segment and a crossing near K along the K-$\Gamma$ segment. This is true in all of the plotted band structures. Both features are present at equilibrium and become exaggerated in the driven system. In Fig.~\ref{bands} we see that bands beyond the first and second valence bands may also exhibit avoided crossings, which are an emergent property of massless Dirac fermions subjected to an antidot potential. Because of the particle-hole symmetry, the same argument applies to the conduction bands as well.

Lastly, we emphasize that the behavior of the two bands closest to zero energy (the inner bands) is consistent with the evolution of the Floquet electronic phases in electric field amplitude from our previous paper (see Fig. 4 in \cite{cupo2021floquet}). The rescaling of the driving parameters required to produce the same phase is partially explained by the slight differences in hole shape and size between the previous and current studies.


\subsection{Optical Conductivity of GAL Floquet Phases}

The homodyne optical conductivity sweeps in the probe frequency corresponding to the discussed quasienergy band structures are visualized in the middle (panels d, e, and f) and right (panels g, h, and i) columns in Figs.~\ref{oc1}-\ref{oc3} \footnote{Note that all plotted conductivities in this paper are normalized to the number of spin channels per band (two)}. In particular, the middle column contains the real parts, while the right column displays the imaginary parts of the optical conductivity. The horizontal axis is the ratio of the probe to the driving angular frequency, $\omega/\Omega$. For a reflectance experiment, an appropriate minimum value of $\omega/\Omega$ is 0.1, and the maximum value is set to 1 since values beyond this would require consideration of the third and fourth order valence/conduction bands that we have neglected \cite{yoo2022spectroscopic}. Each of the plots contains two curves, the longitudinal (blue) and Hall (red) components of the $2 \times 2$ tensor. Notably, in all cases we found numerically that 
\begin{equation}
\sigma_0^{(xx)}(\omega) = \sigma_0^{(yy)}(\omega), \quad \sigma_0^{(yx)}(\omega) = -\sigma_0^{(xy)}(\omega).
\label{longtranscond}
\end{equation}
The equivalence of the longitudinal components is consistent with weak anisotropy in the physical structure of the antidot lattice. Additionally, when anisotropic segments exist in the quasienergy band structures, their contributions to the longitudinal optical conductivity are overshadowed by the presence of strong localization, which occurs for the Floquet semi-Dirac phase, for instance. The relationship between the Hall components agrees with the basic structure of Eq.~\ref{cond} under exchange of $x$ and $y$ for small probe frequencies, provided that the inverse effective mass contribution is negligible.

Let us first focus on the evolution of the real part of the longitudinal conductivity as the driving amplitude increases. We see at equilibrium in Fig.~\ref{oc1}d that there is a single very wide peak, which is consistent with allowed transitions across the entire band structure and is in agreement with previous work \cite{pedersen2008optical}. As the driving amplitude increases in going to Fig.~\ref{oc1}e, the wide peak broadens towards smaller $\omega/\Omega$ values, in accordance with the energy gap decreasing at $\Gamma$. As the gap continues to decrease, a satellite peak emerges just above $\omega/\Omega =$ 0.1 in Fig.~\ref{oc1}f, and its position matches the gap energy.

Upon complete closure of the energy gap, the Floquet Dirac phase appears, and the general behavior of the conductivity is quite different. There are now two distinct peaks located at $\omega/\Omega =$ 0.41 and 0.74 in Fig.~\ref{oc2}d, which equal the inner and outer energy gaps at M. The peak at $\omega/\Omega =$ 0.74 can also be explained by the inner gap at K. In the subsequent plot, Fig.~\ref{oc2}e, the inner gap decreasing at M shifts the first peak to $\omega/\Omega =$ 0.35 and the second peak at $\omega/\Omega =$ 0.81 is close to the outer gap at $\Gamma$. In Fig.~\ref{oc2}f, the two peaks are now located at $\omega/\Omega =$ 0.46 and 0.91, with the lower peak matching the inner gap at $\Gamma$ and the higher peak consistent with the outer gap at both $\Gamma$ and K. 

When the gap closes again and the Floquet semi-Dirac phase is formed, only one peak is visible in Fig.~\ref{oc3}d and this behavior persists in Figs.~\ref{oc3}e and~\ref{oc3}f. The peaks are located at $\omega/\Omega =$ 0.56 (d), 0.67 (e), and 0.75 (f), and all correspond to the inner transition at $\Gamma$.

Altogether, we can summarize the properties of the real part of the longitudinal optical conductivity within the probe frequency range under consideration as follows. For the Floquet quasi-equilibrium phase regime there is a single wide peak, for the Floquet Dirac phase regime there are two relatively narrow peaks, and for the Floquet semi-Dirac phase regime there is one narrow peak. In the above text, we quantified the positions of the peaks and identified plausible corresponding quasienergy band features where possible. Most importantly, each optical conductivity sweep constitutes a unique experimental signature of each Floquet electronic phase. We note that associating peaks in the conductivity with particular band features is only approximate since, in Eq.~\ref{cond}, all k-points in the first BZ contribute to the response at every probe angular frequency $\omega$, the parameter $\eta$ causes broadening, and we only analyze the $m=0$ transitions. In future work, an analysis of selection rules based on the Floquet matrix elements in Eq.~\ref{floquetmatel} may provide further intuition as to why certain transitions dominate the optical conductivity signal \cite{pohle2018symmetry, wang2022polarization, neufeld2019floquet, engelhardt2021dynamical, wang2021observation}.

In principle, the same kind of analysis could be applied to the imaginary part of the longitudinal and real/imaginary parts of the Hall optical conductivity. However, by visual inspection the most dominant features between the four components are related, which is not surprising since: (i) the denominators of the optical conductivity formula Eq.~\ref{cond} that determine the energy resonances are the same for the real/imaginary parts of the longitudinal/Hall components (the difference between them comes from the Floquet matrix elements); and (ii) the real and imaginary components are related to one another by the Kramers-Kronig transformation. Focusing on the dominant features is reasonable, since ultimately one expects to compare theoretical calculations and experimental data, both of which are subject to errors that would likely obscure any fine structure.


\subsection{Properties with no Equilibrium Analog}

Floquet phases are intrinsically non-equilibrium phases, that may display properties with {\em no equilibrium analog}. In particular, for the optical conductivity of the FGAL we observe the following remarkable features:

(i) At equilibrium, the real and imaginary parts of the Hall response are exactly zero, whereas for the driven system they are mostly non-zero and can be as large as the corresponding longitudinal components. 

(ii) Upon the onset of SDL starting in Fig.~\ref{oc2}c, the optical conductivity tends to decrease in magnitude and features soften with each subsequent Floquet phase. 

(iii) We also notice that while the imaginary part of the longitudinal conductivity has one zero at equilibrium, the Floquet phases can have as many as three zeros, see Fig.~\ref{oc2}h for instance. 

However, what is arguably the most surprising and exotic transport property of the FGAL is an anomaly in its Ohmic behavior: While at equilibrium the real parts of the conductivity are greater than or equal to zero, in several Floquet phases these functions can be negative for different ranges of probe frequencies. The importance of this result becomes apparent from the time-averaged power per area expended due to the probe field. Specifically, we have (see Appendix~\ref{sec:joule}),
\begin{equation}
\overline{\frac{dP}{dS}}
=
\frac{1}{2}
\textrm{Re}\{ \sigma_0^{(L)}(\omega) \}
E_{0,\textrm{pro}}^2,
\label{jouledef}
\end{equation}
where $\sigma_0^{(L)}(\omega) \equiv \sigma_0^{(xx)}(\omega) = \sigma_0^{(yy)}(\omega)$. The sign of the real part of the longitudinal conductivity and the time-averaged power per area are the same. Under normal conditions both are positive, which corresponds to the probe-induced current losing energy to the material in the form of local heating. On the other hand, when both quantities are negative the power of the probe is amplified (gain). Intuitively, the material acts as a mediator of power transfer from the Floquet drive to the probe. Note that a negative real part of the longitudinal conductivity has also been observed for several other Floquet quantum materials \cite{chen2018floquet, kumar2020linear, broers2021observing, dabiri2022floquet} and was attributed to the $m \neq 0$ terms in Eq.~\ref{cond} that correspond to the absorption of drive photons \cite{chen2018floquet, dabiri2022floquet}. However, as we emphasized previously, the FGAL features several distinct practical advantages over other Floquet systems.

We distinguish our observation of absolute negative conductivity (ANC) from the more typical negative differential resistance, where the slope of the current-voltage curve can be negative as a result of non-linearity. In semiconductor superlattices \cite{cannon2000absolute} and 2D electron systems \cite{ryzhii2003absolute, inarrea2006zero}, ANC was predicted to occur due to irradiation in the presence of scattering. Experimentally, generating a population inversion in graphene yielded ANC \cite{li2012femtosecond}, which was in agreement with a prior theoretical prediction \cite{ryzhii2007negative}. For doped graphene in the hydrodynamic regime, electron whirlpools near the current-injection contacts result in \textit{local} ANC \cite{bandurin2016negative, levitov2016electron}. These scenarios are to be contrasted with the FGAL, which only requires the generation of Floquet-Bloch bands in an effective single-particle picture to obtain ANC. That is, scattering, population inversion, and non-linearity are {\em not} necessary. In other contexts, effective ANC is produced by parametric amplifiers \cite{decroly1973parametric}, and negative refractive index metamaterials have been studied in great detail \cite{smith2004metamaterials, padilla2006negative, shalaev2007optical, kadic20193d}.

To determine how all of the discussed exotic properties affect the reflectance will require explicitly evaluating a modified version of the standard Fresnel equations to be compatible with a general $2 \times 2$ conductivity tensor \cite{earl2021coherent, michael2022generalized, michael2022fresnel}. We defer this calculation to future work, since the contrast reflectance will depend specifically on the dielectric properties of the supporting substrate, as well as the polarization and angle of incidence of the probe.


\subsection{Remarks on Preparation of Floquet Phases}

In our analysis thus far, we have applied Floquet theory to the idealized situation of perfect invariance under discrete time translation, in which case the physical Hamiltonian defining the FGAL is forever periodic in time and consideration of stroboscopic evolution driven by a time-independent Floquet Hamiltonian suffices. In reality, of course, a Floquet electronic phase of interest must be established by turning on the drive over a finite timescale, with the system initially in equilibrium. Thus, it becomes essential to understand under what timing constraints conclusions derived from Floquet theory are expected to be relevant for practical settings. Issues related to preparation and observation of Floquet phases have been discussed extensively in the literature, requiring in general the use of Floquet non-stroboscopic evolution \cite{bukov2014, bukov2015universal} in order to evaluate the dynamics at intermediate times within a driving period, as well as explicit consideration of quantum control protocols for launching the modulation \cite{goldman2014,rubio2022OCT}.

Previous findings from Ref.\,\cite{d2015dynamical} may be especially useful in our context, since Floquet states are compared quantitatively to time-evolved wave functions obtained from exact quantum dynamics. More precisely, the authors considered a graphene-like material driven by circularly polarized light. The exact time evolution of the system was simulated based on a linear ramping of the vector potential amplitude, assumed to occur over a timescale $\tau$. As a fidelity metric of performance, the squared modulus of the overlap between the exact wave function at the end of the ramping and the target ideal Floquet state was considered. Such a fidelity was numerically determined for several different values of $\tau$, and was found to approach 1 when $\tau$ becomes on the order of about $10^3$ driving cycles. Assuming a similar protocol with our choice of photon energy $\hbar \Omega =$ 0.85 eV, $10^3$ driving cycles would correspond to a total ramping time $\tau \approx$ 5 picoseconds. Thus, while a complete analysis is beyond our scope here and may be more meaningfully carried out with reference to a specific implementation setting, we expect our predictions for the FGAL to remain valid so long as the optical conductivity measurements are performed after the external electric field has been ramped on for several picoseconds.


\section{Conclusions and Outlook}
\label{sec:conc}

In this paper we demonstrated theoretically that the optical conductivity constitutes an experimentally accessible quantity for distinguishing between different Floquet electronic phases. As an example, we considered the Floquet graphene antidot lattice, which we introduced in \cite{cupo2021floquet}. The real and imaginary parts of the longitudinal and Hall components of the homodyne optical conductivity were computed over a wide probe frequency range for a number of quasienergy band structures, which were separated into three phase regimes based on the closing of the energy gap at the Fermi energy. For the real part of the longitudinal conductivity, we predict a single wide peak for the Floquet quasi-equilibrium phase regime, two relatively narrow peaks for the Floquet Dirac phase regime, and one narrow peak for the Floquet semi-Dirac phase regime. The quantitative positions of the peaks further distinguish between different quasienergy band structures within each phase regime.

A number of properties emerged in the Floquet system that \textit{had no analog} in the equilibrium graphene antidot lattice. Notably, the real parts of the longitudinal conductivity can become negative for various ranges of probe frequencies. Physically, the material mediates a transfer of power from the Floquet drive to the probe (gain). Also, while the Hall response is exactly zero at equilibrium, it is generally non-zero and can be as large as the longitudinal conductivity in the Floquet system. Lastly, the flattening of the bands near the superlattice $\Gamma$ point (selective dynamical localization) tends to reduce the overall magnitude of the optical conductivity until, for the final quasienergy band structure plotted, the response is nearly flat compared to equilibrium.

Our work points to several directions for future investigation. While our results are strongly suggestive of a mapping between the Floquet phases and the optical conductivity, additional experimental characterization methods like TR-ARPES may also be considered and numerically investigated \cite{fregoso2013driven}. At the current level of theory, the parameter $\eta$ that represents the electron linewidth is assumed to take an effective constant value. Improvement is possible by using many-body techniques to determine how $\eta$ varies by band, wave vector, and the driving parameters (polarization, amplitude, and frequency) \cite{park2009first}. Also, if the external drive is turned on faster than we discussed previously, i.e., for a quench \cite{dehghani2015out, dehghani2015optical, kumar2020linear, d2015dynamical}, a non-ideal occupation of the Floquet modes should be considered. Furthermore, as mentioned, it would be interesting to use a modified version of the textbook Fresnel equations for a general conductivity tensor in 2D to compute the reflectance explicitly, which may reveal additional features with no equilibrium analog \cite{earl2021coherent, michael2022generalized, michael2022fresnel}. Lastly, while we focused on the optical response in this paper, the homodyne DC Hall conductivity will be an indicator for topological transitions in our system \cite{oka2009photovoltaic, dehghani2015out}, which could occur when the quasienergy gap closes and reopens at the Floquet Dirac or Floquet semi-Dirac phases.


\section*{Acknowledgments}

It is a pleasure to thank Rufus Boyack and Takashi Oka for stimulating discussions. This work was supported by the NSF under grant No. OIA-1921199. The computations in this work were performed on the Discovery cluster and HPC environments supported by the Research Computing group, IT\&C at Dartmouth College. 


\appendix


\section{Computational Details}
\label{sec:comp}

Once the atomic structure of the GAL is defined, the matrix elements of the equilibrium Hamiltonian $\big( \mathcal{H}^{(0)}_{\boldsymbol k} \big)_{uv}$ are found via Eq.~\ref{tbdiag}. The diagonalization in Eq.~\ref{tbsolve} then yields the equilibrium band energies $E_{n \boldsymbol k}$ and eigenvectors $\varphi_{n \boldsymbol k}$. Using Eqs.~\ref{peierls}-\ref{vectorpot}, the Peierls substitution produces the full time-dependent Hamiltonian $\mathcal{H}_{\boldsymbol k}(t)$. Analytical expressions for the full linear electronic current operator $\mathcal{J}_{\boldsymbol k}^{(\alpha)}(t)$ (Eq.~\ref{fullcurrent}) and the full inverse effective mass tensor operator $\mathcal{M}_{\boldsymbol k}^{(\alpha \beta)}(t)$ (Eq.~\ref{fullmass}) are obtained based on Eqs.~\ref{tbdiag} and~\ref{peierls}-\ref{vectorpot}, and then are evaluated numerically. $\mathcal{H}_{\boldsymbol k}(t)$, $\mathcal{J}_{\boldsymbol k}^{(\alpha)}(t)$, and $\mathcal{M}_{\boldsymbol k}^{(\alpha \beta)}(t)$ are computed at every point over the driving cycle, which is sampled uniformly with 100 unique time-points. At this stage, the number of bands for our system is the same as the number of atoms in the unit cell (870). In order to make the subsequent calculations feasible, we dimensionally restrict these three operators to the two bands below and two bands above the Fermi energy (0 eV), see Eqs.~\ref{hamred}-\ref{massred}. With the reduced time-dependent Hamiltonian $H_{\boldsymbol k}(t)$, the block matrix elements $\tilde H_{\boldsymbol k}^{(m,m')}$ in Eq.~\ref{floquetdiag} are evaluated by performing the time-integration numerically. As indicated in the main text, $(m,m')\in[-7,7]$ converges the quasienergies $\epsilon_{n \boldsymbol k}$ from the diagonalization problem in Eq.~\ref{floquetsolve}. We use the eigenvectors $\phi_{n \boldsymbol k}^{(m)}$ to sum the Fourier series in Eq.~\ref{phifourier}, yielding the Floquet eigenstates $\Phi_{n \boldsymbol k}(t)$. \textit{All steps up to this point must be carried out for every wave vector $\boldsymbol k$ under consideration}. 

At this stage, we are equipped to compute the optical conductivity $\sigma_N^{(\alpha \beta)}(\omega)$ via Eq.~\ref{cond}, which contains four nested summations. This will involve performing the numerical time-integration for each of the required Floquet matrix elements, see Eq.~\ref{floquetmatel}. We consider separately the $\alpha \beta = xx, yy, xy, yx$ components of the homodyne ($N = 0$) optical conductivity. Evaluation of Eq.~\ref{cond} is repeated for every distinct value of the probe frequency $\omega$. \textit{The entire procedure of this section is carried out for each of the nine Floquet electronic phases}.

We emphasize that one is not required to use the unfolded quasienergies and the corresponding Floquet eigenstates, but can also take the equivalent quantities within the first Floquet BZ. However, if the photon energy $\hbar \Omega$ is small compared to the natural energy scales of the original system, the automatic shifting and renormalization of the equilibrium bands into the first Floquet BZ significantly complicates the process of determining the proper occupation of the Floquet eigenstates. 

\section{DC Conductivity}
\label{sec:dc}

In this paper we have calculated the optical conductivity for non-zero probe frequencies based on Eq.~\ref{cond}, which can be related to quantities accessible in reflectance experiments. Since connecting a material in a circuit and measuring the current is fundamentally different, we defer DC conductivity calculations to future work. Nevertheless, for the reader interested in computing these properties, the DC conductivity is obtained by taking the limit as $\omega \rightarrow 0$ of Eq.~\ref{cond}, yielding \cite{kumar2020linear}
\begin{equation}
\begin{split}
& \sigma_N^{(\alpha \beta)}(0)
=
-i \hbar
\int_{\textrm{BZ}}^{} \frac{d^\mathcal{D} k}{(2 \pi)^{\mathcal{D}}}
\sum_{n}^{} 
f_{n \boldsymbol k}
\\
& \sum_{n' (n' \neq n)}^{}
\sum_{m}^{}
\frac{1}{( \epsilon_{n \boldsymbol k} - \epsilon_{n' \boldsymbol k} - m \hbar \Omega )^2}
\\
& \Big\{
\mathcal{F}_{(n,-N;n',m)\boldsymbol k}[j_{\boldsymbol k}^{(\alpha)}(t)]
\mathcal{F}_{(n',m;n,0)\boldsymbol k}[j_{\boldsymbol k}^{(\beta)}(t)]
\\
& -
\mathcal{F}_{(n,0;n',m)\boldsymbol k}[j_{\boldsymbol k}^{(\beta)}(t)]
\mathcal{F}_{(n',m;n,N)\boldsymbol k}[j_{\boldsymbol k}^{(\alpha)}(t)]
\Big\}.
\end{split}
\label{conddc}
\end{equation}
In 2D a Floquet analog of the Thouless-Kohmoto-Nightingale-den Nijs (TKNN) relation is found, namely, 
\begin{equation}
\sigma_0^{(xy)}(0)
=
\frac{e^2}{h}
\sum_{n}^{}
f_{n}
C_n ,
\label{tknn}
\end{equation}
with the Chern number
\begin{equation}
C_n
\equiv 
\frac{i}{2 \pi}
\int_{\textrm{BZ}}^{} d^2 k 
F_{n \boldsymbol k}^{(0)}.
\label{chern}
\end{equation}
The Berry curvature is time-periodic,
\begin{equation}
\begin{split}
& F_{n \boldsymbol k}(t) = \sum_{m}^{} F_{n \boldsymbol k}^{(m)} e^{-i m \Omega t}
\\
& = \bigg(
\frac{\partial \Phi^\dagger_{n \boldsymbol k}(t)}{\partial k_x} \frac{\partial \Phi_{n \boldsymbol k}(t)}{\partial k_y}
-
\frac{\partial \Phi^\dagger_{n \boldsymbol k}(t)}{\partial k_y} \frac{\partial \Phi_{n \boldsymbol k}(t)}{\partial k_x}
\bigg),
\end{split}
\label{berryfourier}
\end{equation}
where the Chern number calculation utilizes the static component
\begin{equation}
F_{n \boldsymbol k}^{(0)}
=
\frac{1}{T}
\int_{0}^{T}
dt\,
F_{n \boldsymbol k}(t).
\label{berry}
\end{equation}
The homodyne DC Hall conductivity is quantized for ideal occupation ($f_{n \boldsymbol k}$ is only 0 or 1) and is an indicator for topological transitions.


\onecolumngrid
\vspace*{5mm}

\section{Probe Power per Area}
\label{sec:joule}

\noindent The power per area expended due to the probe field is given by 
\begin{equation}
\frac{dP(t)}{dS}
=
\delta \left\langle \boldsymbol J \right\rangle_\textrm{physical}(t)
\cdot
\boldsymbol E_{\textrm{pro,physical}}(t).
\label{joule_appendix}
\end{equation}
Using Eqs.~\ref{realprobe} and \ref{realcurrent}, this expands to
$$
\frac{dP(t)}{dS} 
= 
\sum_{N}^{} \Big[ \textrm{Re}\{\sigma_N(\omega)\} \textrm{cos}((\omega + N \Omega) t) + \textrm{Im}\{\sigma_N(\omega)\} \textrm{sin}((\omega + N \Omega) t) \Big] \boldsymbol E_{0,\textrm{pro}}
\cdot
\textrm{cos}(\omega t) \boldsymbol E_{0,\textrm{pro}}.
$$
As we only consider the homodyne ($N = 0$) contribution to the total conductivity, we have the following simplification:
$$
\frac{dP(t)}{dS} 
= 
\bigg[ \textrm{Re}\{\sigma_0(\omega)\} \textrm{cos}^2(\omega t) + \frac{1}{2} \textrm{Im}\{\sigma_0(\omega)\} \textrm{sin}(2 \omega t) \bigg] \boldsymbol E_{0,\textrm{pro}}
\cdot
\boldsymbol E_{0,\textrm{pro}}.
$$
To obtain the time-averaged power per area, we use the standard time-averages $\overline{\textrm{cos}^2(\omega t)} = 0.5$ and $\overline{\textrm{sin}(2 \omega t)} = 0$, resulting in
$$
\overline{\frac{dP}{dS}}
=
\frac{1}{2} \textrm{Re}\{\sigma_0(\omega)\}  \boldsymbol E_{0,\textrm{pro}}
\cdot
\boldsymbol E_{0,\textrm{pro}}.
$$
The conductivity tensor has the form (see Eq.~\ref{longtranscond})
$$
\sigma_0(\omega)
=
\begin{bmatrix}
\sigma_0^{(xx)}(\omega) & \sigma_0^{(xy)}(\omega) \\
\sigma_0^{(yx)}(\omega) & \sigma_0^{(yy)}(\omega)
\end{bmatrix}
=
\begin{bmatrix}
\sigma_0^{(xx)}(\omega) & \sigma_0^{(xy)}(\omega) \\
-\sigma_0^{(xy)}(\omega) & \sigma_0^{(xx)}(\omega)
\end{bmatrix}
\equiv
\begin{bmatrix}
\sigma_0^{(L)}(\omega) & \sigma_0^{(T)}(\omega) \\
-\sigma_0^{(T)}(\omega) & \sigma_0^{(L)}(\omega)
\end{bmatrix}
.
$$
For a linearly polarized probe we have
$$
\tilde{\boldsymbol E}_{0,\textrm{pro}}
=
(N_x \hat{x} + N_y \hat{y} + N_z \hat{z})
\tilde{E}_{0,\textrm{pro}}
\ , \quad
N_x^2 + N_y^2 + N_z^2 = 1
.
$$
Since we consider a 2D material in the $z=0$ plane, only the $xy$ projection will contribute to the response:
$$
\boldsymbol E_{0,\textrm{pro}}
=
(N_x \hat{x} + N_y \hat{y})
\tilde{E}_{0,\textrm{pro}},
$$
$$
\boldsymbol E_{0,\textrm{pro}}
=
\Bigg(
\frac{N_x}{\sqrt{N_x^2 + N_y^2}}
\hat{x}
+
\frac{N_y}{\sqrt{N_x^2 + N_y^2}}
\hat{y}
\Bigg)
\Big( \sqrt{N_x^2 + N_y^2} \tilde{E}_{0,\textrm{pro}} \Big)
,
$$
$$
\boldsymbol E_{0,\textrm{pro}}
\equiv
(n_x \hat{x} + n_y \hat{y})
E_{0,\textrm{pro}}
\ , \quad
n_x^2 + n_y^2 = 1
.
$$
The time-averaged power per area then expands to
$$
\overline{\frac{dP}{dS}}
=
\frac{1}{2}
\begin{bmatrix}
\textrm{Re}\{\sigma_0^{(L)}(\omega)\} & \textrm{Re}\{\sigma_0^{(T)}(\omega)\} \\
-\textrm{Re}\{\sigma_0^{(T)}(\omega)\} & \textrm{Re}\{\sigma_0^{(L)}(\omega)\}
\end{bmatrix}
(n_x \hat{x} + n_y \hat{y})
E_{0,\textrm{pro}}
\cdot
(n_x \hat{x} + n_y \hat{y})
E_{0,\textrm{pro}}.
$$
The contributions from the transverse components of the conductivity cancel and we are simply left with
\begin{equation}
\overline{\frac{dP}{dS}}
=
\frac{1}{2}
\textrm{Re}\{ \sigma_0^{(L)}(\omega) \}
E_{0,\textrm{pro}}^2,
\label{jouledef_appendix}
\end{equation}
which is Eq.~\ref{jouledef} in the main text.

\twocolumngrid


\bibliography{references}

\begin{thebibliography}{80}%
\makeatletter
\providecommand \@ifxundefined [1]{%
 \@ifx{#1\undefined}
}%
\providecommand \@ifnum [1]{%
 \ifnum #1\expandafter \@firstoftwo
 \else \expandafter \@secondoftwo
 \fi
}%
\providecommand \@ifx [1]{%
 \ifx #1\expandafter \@firstoftwo
 \else \expandafter \@secondoftwo
 \fi
}%
\providecommand \natexlab [1]{#1}%
\providecommand \enquote  [1]{``#1''}%
\providecommand \bibnamefont  [1]{#1}%
\providecommand \bibfnamefont [1]{#1}%
\providecommand \citenamefont [1]{#1}%
\providecommand \href@noop [0]{\@secondoftwo}%
\providecommand \href [0]{\begingroup \@sanitize@url \@href}%
\providecommand \@href[1]{\@@startlink{#1}\@@href}%
\providecommand \@@href[1]{\endgroup#1\@@endlink}%
\providecommand \@sanitize@url [0]{\catcode `\\12\catcode `\$12\catcode
  `\&12\catcode `\#12\catcode `\^12\catcode `\_12\catcode `\%12\relax}%
\providecommand \@@startlink[1]{}%
\providecommand \@@endlink[0]{}%
\providecommand \url  [0]{\begingroup\@sanitize@url \@url }%
\providecommand \@url [1]{\endgroup\@href {#1}{\urlprefix }}%
\providecommand \urlprefix  [0]{URL }%
\providecommand \Eprint [0]{\href }%
\providecommand \doibase [0]{https://doi.org/}%
\providecommand \selectlanguage [0]{\@gobble}%
\providecommand \bibinfo  [0]{\@secondoftwo}%
\providecommand \bibfield  [0]{\@secondoftwo}%
\providecommand \translation [1]{[#1]}%
\providecommand \BibitemOpen [0]{}%
\providecommand \bibitemStop [0]{}%
\providecommand \bibitemNoStop [0]{.\EOS\space}%
\providecommand \EOS [0]{\spacefactor3000\relax}%
\providecommand \BibitemShut  [1]{\csname bibitem#1\endcsname}%
\let\auto@bib@innerbib\@empty
\bibitem [{\citenamefont {Bukov}\ \emph {et~al.}(2015)\citenamefont {Bukov},
  \citenamefont {D'Alessio},\ and\ \citenamefont
  {Polkovnikov}}]{bukov2015universal}%
  \BibitemOpen
  \bibfield  {author} {\bibinfo {author} {\bibfnamefont {M.}~\bibnamefont
  {Bukov}}, \bibinfo {author} {\bibfnamefont {L.}~\bibnamefont {D'Alessio}},\
  and\ \bibinfo {author} {\bibfnamefont {A.}~\bibnamefont {Polkovnikov}},\
  }\bibfield  {title} {\bibinfo {title} {Universal high-frequency behavior of
  periodically driven systems: {F}rom dynamical stabilization to {F}loquet
  engineering},\ }\href@noop {} {\bibfield  {journal} {\bibinfo  {journal}
  {Adv. Phys.}\ }\textbf {\bibinfo {volume} {64}},\ \bibinfo {pages} {139}
  (\bibinfo {year} {2015})}\BibitemShut {NoStop}%
\bibitem [{\citenamefont {Basov}\ \emph {et~al.}(2017)\citenamefont {Basov},
  \citenamefont {Averitt},\ and\ \citenamefont {Hsieh}}]{basov2017towards}%
  \BibitemOpen
  \bibfield  {author} {\bibinfo {author} {\bibfnamefont {D.~N.}\ \bibnamefont
  {Basov}}, \bibinfo {author} {\bibfnamefont {R.~D.}\ \bibnamefont {Averitt}},\
  and\ \bibinfo {author} {\bibfnamefont {D.}~\bibnamefont {Hsieh}},\ }\bibfield
   {title} {\bibinfo {title} {Towards properties on demand in quantum
  materials},\ }\href@noop {} {\bibfield  {journal} {\bibinfo  {journal} {Nat.
  Mater.}\ }\textbf {\bibinfo {volume} {16}},\ \bibinfo {pages} {1077}
  (\bibinfo {year} {2017})}\BibitemShut {NoStop}%
\bibitem [{\citenamefont {Oka}\ and\ \citenamefont
  {Kitamura}(2019)}]{oka2019floquet}%
  \BibitemOpen
  \bibfield  {author} {\bibinfo {author} {\bibfnamefont {T.}~\bibnamefont
  {Oka}}\ and\ \bibinfo {author} {\bibfnamefont {S.}~\bibnamefont {Kitamura}},\
  }\bibfield  {title} {\bibinfo {title} {Floquet engineering of quantum
  materials},\ }\href@noop {} {\bibfield  {journal} {\bibinfo  {journal} {Annu.
  Rev. Condens. Matter Phys.}\ }\textbf {\bibinfo {volume} {10}},\ \bibinfo
  {pages} {387} (\bibinfo {year} {2019})}\BibitemShut {NoStop}%
\bibitem [{\citenamefont {De~Giovannini}\ and\ \citenamefont
  {H{\"u}bener}(2020)}]{de2019floquet}%
  \BibitemOpen
  \bibfield  {author} {\bibinfo {author} {\bibfnamefont {U.}~\bibnamefont
  {De~Giovannini}}\ and\ \bibinfo {author} {\bibfnamefont {H.}~\bibnamefont
  {H{\"u}bener}},\ }\bibfield  {title} {\bibinfo {title} {Floquet analysis of
  excitations in materials},\ }\href@noop {} {\bibfield  {journal} {\bibinfo
  {journal} {J. Phys. Mater.}\ }\textbf {\bibinfo {volume} {3}},\ \bibinfo
  {pages} {012001} (\bibinfo {year} {2020})}\BibitemShut {NoStop}%
\bibitem [{\citenamefont {Harper}\ \emph {et~al.}(2020)\citenamefont {Harper},
  \citenamefont {Roy}, \citenamefont {Rudner},\ and\ \citenamefont
  {Sondhi}}]{harper2020topology}%
  \BibitemOpen
  \bibfield  {author} {\bibinfo {author} {\bibfnamefont {F.}~\bibnamefont
  {Harper}}, \bibinfo {author} {\bibfnamefont {R.}~\bibnamefont {Roy}},
  \bibinfo {author} {\bibfnamefont {M.~S.}\ \bibnamefont {Rudner}},\ and\
  \bibinfo {author} {\bibfnamefont {S.~L.}\ \bibnamefont {Sondhi}},\ }\bibfield
   {title} {\bibinfo {title} {Topology and broken symmetry in {F}loquet
  systems},\ }\href@noop {} {\bibfield  {journal} {\bibinfo  {journal} {Annu.
  Rev. Condens. Matter Phys.}\ }\textbf {\bibinfo {volume} {11}},\ \bibinfo
  {pages} {345} (\bibinfo {year} {2020})}\BibitemShut {NoStop}%
\bibitem [{\citenamefont {Rudner}\ and\ \citenamefont
  {Lindner}(2020{\natexlab{a}})}]{rudner2020band}%
  \BibitemOpen
  \bibfield  {author} {\bibinfo {author} {\bibfnamefont {M.~S.}\ \bibnamefont
  {Rudner}}\ and\ \bibinfo {author} {\bibfnamefont {N.~H.}\ \bibnamefont
  {Lindner}},\ }\bibfield  {title} {\bibinfo {title} {Band structure
  engineering and non-equilibrium dynamics in {F}loquet topological
  insulators},\ }\href@noop {} {\bibfield  {journal} {\bibinfo  {journal} {Nat.
  Rev. Phys.}\ }\textbf {\bibinfo {volume} {2}},\ \bibinfo {pages} {229}
  (\bibinfo {year} {2020}{\natexlab{a}})}\BibitemShut {NoStop}%
\bibitem [{\citenamefont {Rodriguez-Vega}\ \emph {et~al.}(2021)\citenamefont
  {Rodriguez-Vega}, \citenamefont {Vogl},\ and\ \citenamefont
  {Fiete}}]{rodriguez2021low}%
  \BibitemOpen
  \bibfield  {author} {\bibinfo {author} {\bibfnamefont {M.}~\bibnamefont
  {Rodriguez-Vega}}, \bibinfo {author} {\bibfnamefont {M.}~\bibnamefont
  {Vogl}},\ and\ \bibinfo {author} {\bibfnamefont {G.~A.}\ \bibnamefont
  {Fiete}},\ }\bibfield  {title} {\bibinfo {title} {Low-frequency and
  {M}oir{\'e}--{F}loquet engineering: {A} review},\ }\href@noop {} {\bibfield
  {journal} {\bibinfo  {journal} {Ann. Phys.}\ }\textbf {\bibinfo {volume}
  {435}},\ \bibinfo {pages} {168434} (\bibinfo {year} {2021})}\BibitemShut
  {NoStop}%
\bibitem [{\citenamefont {Poudel}\ \emph {et~al.}(2015)\citenamefont {Poudel},
  \citenamefont {Ortiz},\ and\ \citenamefont {Viola}}]{Poudel}%
  \BibitemOpen
  \bibfield  {author} {\bibinfo {author} {\bibfnamefont {A.}~\bibnamefont
  {Poudel}}, \bibinfo {author} {\bibfnamefont {G.}~\bibnamefont {Ortiz}},\ and\
  \bibinfo {author} {\bibfnamefont {L.}~\bibnamefont {Viola}},\ }\bibfield
  {title} {\bibinfo {title} {Dynamical generation of {F}loquet {M}ajorana flat
  bands in $s$-wave superconductors},\ }\href@noop {} {\bibfield  {journal}
  {\bibinfo  {journal} {EPL}\ }\textbf {\bibinfo {volume} {110}},\ \bibinfo
  {pages} {17004} (\bibinfo {year} {2015})}\BibitemShut {NoStop}%
\bibitem [{\citenamefont {Shirley}(1965)}]{shirley1965solution}%
  \BibitemOpen
  \bibfield  {author} {\bibinfo {author} {\bibfnamefont {J.~H.}\ \bibnamefont
  {Shirley}},\ }\bibfield  {title} {\bibinfo {title} {Solution of the
  {S}chr{\"o}dinger equation with a {H}amiltonian periodic in time},\
  }\href@noop {} {\bibfield  {journal} {\bibinfo  {journal} {Phys. Rev.}\
  }\textbf {\bibinfo {volume} {138}},\ \bibinfo {pages} {B979} (\bibinfo {year}
  {1965})}\BibitemShut {NoStop}%
\bibitem [{\citenamefont {Sambe}(1973)}]{sambe1973steady}%
  \BibitemOpen
  \bibfield  {author} {\bibinfo {author} {\bibfnamefont {H.}~\bibnamefont
  {Sambe}},\ }\bibfield  {title} {\bibinfo {title} {Steady states and
  quasienergies of a quantum-mechanical system in an oscillating field},\
  }\href@noop {} {\bibfield  {journal} {\bibinfo  {journal} {Phys. Rev. A}\
  }\textbf {\bibinfo {volume} {7}},\ \bibinfo {pages} {2203} (\bibinfo {year}
  {1973})}\BibitemShut {NoStop}%
\bibitem [{\citenamefont {Rudner}\ and\ \citenamefont
  {Lindner}(2020{\natexlab{b}})}]{rudner2020floquet}%
  \BibitemOpen
  \bibfield  {author} {\bibinfo {author} {\bibfnamefont {M.~S.}\ \bibnamefont
  {Rudner}}\ and\ \bibinfo {author} {\bibfnamefont {N.~H.}\ \bibnamefont
  {Lindner}},\ }\bibfield  {title} {\bibinfo {title} {The {F}loquet engineer's
  handbook},\ }\href@noop {} {\bibfield  {journal} {\bibinfo  {journal}
  {arXiv:2003.08252v2}\ } (\bibinfo {year} {2020}{\natexlab{b}})}\BibitemShut
  {NoStop}%
\bibitem [{\citenamefont {Wang}\ \emph {et~al.}(2013)\citenamefont {Wang},
  \citenamefont {Steinberg}, \citenamefont {Jarillo-Herrero},\ and\
  \citenamefont {Gedik}}]{wang2013observation}%
  \BibitemOpen
  \bibfield  {author} {\bibinfo {author} {\bibfnamefont {Y.~H.}\ \bibnamefont
  {Wang}}, \bibinfo {author} {\bibfnamefont {H.}~\bibnamefont {Steinberg}},
  \bibinfo {author} {\bibfnamefont {P.}~\bibnamefont {Jarillo-Herrero}},\ and\
  \bibinfo {author} {\bibfnamefont {N.}~\bibnamefont {Gedik}},\ }\bibfield
  {title} {\bibinfo {title} {Observation of {F}loquet-{B}loch states on the
  surface of a topological insulator},\ }\href@noop {} {\bibfield  {journal}
  {\bibinfo  {journal} {Science}\ }\textbf {\bibinfo {volume} {342}},\ \bibinfo
  {pages} {453} (\bibinfo {year} {2013})}\BibitemShut {NoStop}%
\bibitem [{\citenamefont {Mahmood}\ \emph {et~al.}(2016)\citenamefont
  {Mahmood}, \citenamefont {Chan}, \citenamefont {Alpichshev}, \citenamefont
  {Gardner}, \citenamefont {Lee}, \citenamefont {Lee},\ and\ \citenamefont
  {Gedik}}]{mahmood2016selective}%
  \BibitemOpen
  \bibfield  {author} {\bibinfo {author} {\bibfnamefont {F.}~\bibnamefont
  {Mahmood}}, \bibinfo {author} {\bibfnamefont {C.-K.}\ \bibnamefont {Chan}},
  \bibinfo {author} {\bibfnamefont {Z.}~\bibnamefont {Alpichshev}}, \bibinfo
  {author} {\bibfnamefont {D.}~\bibnamefont {Gardner}}, \bibinfo {author}
  {\bibfnamefont {Y.}~\bibnamefont {Lee}}, \bibinfo {author} {\bibfnamefont
  {P.~A.}\ \bibnamefont {Lee}},\ and\ \bibinfo {author} {\bibfnamefont
  {N.}~\bibnamefont {Gedik}},\ }\bibfield  {title} {\bibinfo {title} {Selective
  scattering between {F}loquet--{B}loch and {V}olkov states in a topological
  insulator},\ }\href@noop {} {\bibfield  {journal} {\bibinfo  {journal} {Nat.
  Phys.}\ }\textbf {\bibinfo {volume} {12}},\ \bibinfo {pages} {306} (\bibinfo
  {year} {2016})}\BibitemShut {NoStop}%
\bibitem [{\citenamefont {Reimann}\ \emph {et~al.}(2018)\citenamefont
  {Reimann}, \citenamefont {Schlauderer}, \citenamefont {Schmid}, \citenamefont
  {Langer}, \citenamefont {Baierl}, \citenamefont {Kokh}, \citenamefont
  {Tereshchenko}, \citenamefont {Kimura}, \citenamefont {Lange}, \citenamefont
  {G{\"u}dde} \emph {et~al.}}]{reimann2018subcycle}%
  \BibitemOpen
  \bibfield  {author} {\bibinfo {author} {\bibfnamefont {J.}~\bibnamefont
  {Reimann}}, \bibinfo {author} {\bibfnamefont {S.}~\bibnamefont
  {Schlauderer}}, \bibinfo {author} {\bibfnamefont {C.~P.}\ \bibnamefont
  {Schmid}}, \bibinfo {author} {\bibfnamefont {F.}~\bibnamefont {Langer}},
  \bibinfo {author} {\bibfnamefont {S.}~\bibnamefont {Baierl}}, \bibinfo
  {author} {\bibfnamefont {K.~A.}\ \bibnamefont {Kokh}}, \bibinfo {author}
  {\bibfnamefont {O.~E.}\ \bibnamefont {Tereshchenko}}, \bibinfo {author}
  {\bibfnamefont {A.}~\bibnamefont {Kimura}}, \bibinfo {author} {\bibfnamefont
  {C.}~\bibnamefont {Lange}}, \bibinfo {author} {\bibfnamefont
  {J.}~\bibnamefont {G{\"u}dde}}, \emph {et~al.},\ }\bibfield  {title}
  {\bibinfo {title} {Subcycle observation of lightwave-driven {D}irac currents
  in a topological surface band},\ }\href@noop {} {\bibfield  {journal}
  {\bibinfo  {journal} {Nature}\ }\textbf {\bibinfo {volume} {562}},\ \bibinfo
  {pages} {396} (\bibinfo {year} {2018})}\BibitemShut {NoStop}%
\bibitem [{\citenamefont {Reutzel}\ \emph {et~al.}(2020)\citenamefont
  {Reutzel}, \citenamefont {Li}, \citenamefont {Wang},\ and\ \citenamefont
  {Petek}}]{reutzel2020coherent}%
  \BibitemOpen
  \bibfield  {author} {\bibinfo {author} {\bibfnamefont {M.}~\bibnamefont
  {Reutzel}}, \bibinfo {author} {\bibfnamefont {A.}~\bibnamefont {Li}},
  \bibinfo {author} {\bibfnamefont {Z.}~\bibnamefont {Wang}},\ and\ \bibinfo
  {author} {\bibfnamefont {H.}~\bibnamefont {Petek}},\ }\bibfield  {title}
  {\bibinfo {title} {Coherent multidimensional photoelectron spectroscopy of
  ultrafast quasiparticle dressing by light},\ }\href@noop {} {\bibfield
  {journal} {\bibinfo  {journal} {Nat. Commun.}\ }\textbf {\bibinfo {volume}
  {11}},\ \bibinfo {pages} {1} (\bibinfo {year} {2020})}\BibitemShut {NoStop}%
\bibitem [{\citenamefont {Keunecke}\ \emph {et~al.}(2020)\citenamefont
  {Keunecke}, \citenamefont {Reutzel}, \citenamefont {Schmitt}, \citenamefont
  {Osterkorn}, \citenamefont {Mishra}, \citenamefont {M{\"o}ller},
  \citenamefont {Bennecke}, \citenamefont {Matthijs~Jansen}, \citenamefont
  {Steil}, \citenamefont {Manmana} \emph
  {et~al.}}]{keunecke2020electromagnetic}%
  \BibitemOpen
  \bibfield  {author} {\bibinfo {author} {\bibfnamefont {M.}~\bibnamefont
  {Keunecke}}, \bibinfo {author} {\bibfnamefont {M.}~\bibnamefont {Reutzel}},
  \bibinfo {author} {\bibfnamefont {D.}~\bibnamefont {Schmitt}}, \bibinfo
  {author} {\bibfnamefont {A.}~\bibnamefont {Osterkorn}}, \bibinfo {author}
  {\bibfnamefont {T.~A.}\ \bibnamefont {Mishra}}, \bibinfo {author}
  {\bibfnamefont {C.}~\bibnamefont {M{\"o}ller}}, \bibinfo {author}
  {\bibfnamefont {W.}~\bibnamefont {Bennecke}}, \bibinfo {author}
  {\bibfnamefont {G.~S.}\ \bibnamefont {Matthijs~Jansen}}, \bibinfo {author}
  {\bibfnamefont {D.}~\bibnamefont {Steil}}, \bibinfo {author} {\bibfnamefont
  {S.~R.}\ \bibnamefont {Manmana}}, \emph {et~al.},\ }\bibfield  {title}
  {\bibinfo {title} {Electromagnetic dressing of the electron energy spectrum
  of {A}u (111) at high momenta},\ }\href@noop {} {\bibfield  {journal}
  {\bibinfo  {journal} {Phys. Rev. B}\ }\textbf {\bibinfo {volume} {102}},\
  \bibinfo {pages} {161403(R)} (\bibinfo {year} {2020})}\BibitemShut {NoStop}%
\bibitem [{\citenamefont {Aeschlimann}\ \emph {et~al.}(2021)\citenamefont
  {Aeschlimann}, \citenamefont {Sato}, \citenamefont {Krause}, \citenamefont
  {Ch{\'a}vez-Cervantes}, \citenamefont {De~Giovannini}, \citenamefont
  {H{\"u}bener}, \citenamefont {Forti}, \citenamefont {Coletti}, \citenamefont
  {Hanff}, \citenamefont {Rossnagel} \emph {et~al.}}]{aeschlimann2021survival}%
  \BibitemOpen
  \bibfield  {author} {\bibinfo {author} {\bibfnamefont {S.}~\bibnamefont
  {Aeschlimann}}, \bibinfo {author} {\bibfnamefont {S.~A.}\ \bibnamefont
  {Sato}}, \bibinfo {author} {\bibfnamefont {R.}~\bibnamefont {Krause}},
  \bibinfo {author} {\bibfnamefont {M.}~\bibnamefont {Ch{\'a}vez-Cervantes}},
  \bibinfo {author} {\bibfnamefont {U.}~\bibnamefont {De~Giovannini}}, \bibinfo
  {author} {\bibfnamefont {H.}~\bibnamefont {H{\"u}bener}}, \bibinfo {author}
  {\bibfnamefont {S.}~\bibnamefont {Forti}}, \bibinfo {author} {\bibfnamefont
  {C.}~\bibnamefont {Coletti}}, \bibinfo {author} {\bibfnamefont
  {K.}~\bibnamefont {Hanff}}, \bibinfo {author} {\bibfnamefont
  {K.}~\bibnamefont {Rossnagel}}, \emph {et~al.},\ }\bibfield  {title}
  {\bibinfo {title} {Survival of {F}loquet--{B}loch states in the presence of
  scattering},\ }\href@noop {} {\bibfield  {journal} {\bibinfo  {journal} {Nano
  Lett.}\ }\textbf {\bibinfo {volume} {21}},\ \bibinfo {pages} {5028} (\bibinfo
  {year} {2021})}\BibitemShut {NoStop}%
\bibitem [{\citenamefont {Zhou}\ \emph {et~al.}(2023)\citenamefont {Zhou},
  \citenamefont {Bao}, \citenamefont {Fan}, \citenamefont {Zhou}, \citenamefont
  {Gao}, \citenamefont {Zhong}, \citenamefont {Lin}, \citenamefont {Liu},
  \citenamefont {Yu}, \citenamefont {Tang} \emph
  {et~al.}}]{zhou2023pseudospin}%
  \BibitemOpen
  \bibfield  {author} {\bibinfo {author} {\bibfnamefont {S.}~\bibnamefont
  {Zhou}}, \bibinfo {author} {\bibfnamefont {C.}~\bibnamefont {Bao}}, \bibinfo
  {author} {\bibfnamefont {B.}~\bibnamefont {Fan}}, \bibinfo {author}
  {\bibfnamefont {H.}~\bibnamefont {Zhou}}, \bibinfo {author} {\bibfnamefont
  {Q.}~\bibnamefont {Gao}}, \bibinfo {author} {\bibfnamefont {H.}~\bibnamefont
  {Zhong}}, \bibinfo {author} {\bibfnamefont {T.}~\bibnamefont {Lin}}, \bibinfo
  {author} {\bibfnamefont {H.}~\bibnamefont {Liu}}, \bibinfo {author}
  {\bibfnamefont {P.}~\bibnamefont {Yu}}, \bibinfo {author} {\bibfnamefont
  {P.}~\bibnamefont {Tang}}, \emph {et~al.},\ }\bibfield  {title} {\bibinfo
  {title} {Pseudospin-selective {F}loquet band engineering in black
  phosphorus},\ }\href@noop {} {\bibfield  {journal} {\bibinfo  {journal}
  {Nature}\ }\textbf {\bibinfo {volume} {614}},\ \bibinfo {pages} {75}
  (\bibinfo {year} {2023})}\BibitemShut {NoStop}%
\bibitem [{\citenamefont {McIver}\ \emph {et~al.}(2020)\citenamefont {McIver},
  \citenamefont {Schulte}, \citenamefont {Stein}, \citenamefont {Matsuyama},
  \citenamefont {Jotzu}, \citenamefont {Meier},\ and\ \citenamefont
  {Cavalleri}}]{mciver2020light}%
  \BibitemOpen
  \bibfield  {author} {\bibinfo {author} {\bibfnamefont {J.~W.}\ \bibnamefont
  {McIver}}, \bibinfo {author} {\bibfnamefont {B.}~\bibnamefont {Schulte}},
  \bibinfo {author} {\bibfnamefont {F.-U.}\ \bibnamefont {Stein}}, \bibinfo
  {author} {\bibfnamefont {T.}~\bibnamefont {Matsuyama}}, \bibinfo {author}
  {\bibfnamefont {G.}~\bibnamefont {Jotzu}}, \bibinfo {author} {\bibfnamefont
  {G.}~\bibnamefont {Meier}},\ and\ \bibinfo {author} {\bibfnamefont
  {A.}~\bibnamefont {Cavalleri}},\ }\bibfield  {title} {\bibinfo {title}
  {Light-induced anomalous {H}all effect in graphene},\ }\href@noop {}
  {\bibfield  {journal} {\bibinfo  {journal} {Nat. Phys.}\ }\textbf {\bibinfo
  {volume} {16}},\ \bibinfo {pages} {38} (\bibinfo {year} {2020})}\BibitemShut
  {NoStop}%
\bibitem [{\citenamefont {Candussio}\ \emph {et~al.}(2022)\citenamefont
  {Candussio}, \citenamefont {Bernreuter}, \citenamefont {Rockinger},
  \citenamefont {Watanabe}, \citenamefont {Taniguchi}, \citenamefont {Eroms},
  \citenamefont {Dmitriev}, \citenamefont {Weiss},\ and\ \citenamefont
  {Ganichev}}]{candussio2022terahertz}%
  \BibitemOpen
  \bibfield  {author} {\bibinfo {author} {\bibfnamefont {S.}~\bibnamefont
  {Candussio}}, \bibinfo {author} {\bibfnamefont {S.}~\bibnamefont
  {Bernreuter}}, \bibinfo {author} {\bibfnamefont {T.}~\bibnamefont
  {Rockinger}}, \bibinfo {author} {\bibfnamefont {K.}~\bibnamefont {Watanabe}},
  \bibinfo {author} {\bibfnamefont {T.}~\bibnamefont {Taniguchi}}, \bibinfo
  {author} {\bibfnamefont {J.}~\bibnamefont {Eroms}}, \bibinfo {author}
  {\bibfnamefont {I.~A.}\ \bibnamefont {Dmitriev}}, \bibinfo {author}
  {\bibfnamefont {D.}~\bibnamefont {Weiss}},\ and\ \bibinfo {author}
  {\bibfnamefont {S.~D.}\ \bibnamefont {Ganichev}},\ }\bibfield  {title}
  {\bibinfo {title} {Terahertz radiation induced circular {H}all effect in
  graphene},\ }\href@noop {} {\bibfield  {journal} {\bibinfo  {journal} {Phys.
  Rev. B}\ }\textbf {\bibinfo {volume} {105}},\ \bibinfo {pages} {155416}
  (\bibinfo {year} {2022})}\BibitemShut {NoStop}%
\bibitem [{\citenamefont {Fregoso}\ \emph {et~al.}(2013)\citenamefont
  {Fregoso}, \citenamefont {Wang}, \citenamefont {Gedik},\ and\ \citenamefont
  {Galitski}}]{fregoso2013driven}%
  \BibitemOpen
  \bibfield  {author} {\bibinfo {author} {\bibfnamefont {B.~M.}\ \bibnamefont
  {Fregoso}}, \bibinfo {author} {\bibfnamefont {Y.~H.}\ \bibnamefont {Wang}},
  \bibinfo {author} {\bibfnamefont {N.}~\bibnamefont {Gedik}},\ and\ \bibinfo
  {author} {\bibfnamefont {V.}~\bibnamefont {Galitski}},\ }\bibfield  {title}
  {\bibinfo {title} {Driven electronic states at the surface of a topological
  insulator},\ }\href@noop {} {\bibfield  {journal} {\bibinfo  {journal} {Phys.
  Rev. B}\ }\textbf {\bibinfo {volume} {88}},\ \bibinfo {pages} {155129}
  (\bibinfo {year} {2013})}\BibitemShut {NoStop}%
\bibitem [{\citenamefont {Sato}\ \emph {et~al.}(2019)\citenamefont {Sato},
  \citenamefont {McIver}, \citenamefont {Nuske}, \citenamefont {Tang},
  \citenamefont {Jotzu}, \citenamefont {Schulte}, \citenamefont {H{\"u}bener},
  \citenamefont {De~Giovannini}, \citenamefont {Mathey}, \citenamefont {Sentef}
  \emph {et~al.}}]{sato2019microscopic}%
  \BibitemOpen
  \bibfield  {author} {\bibinfo {author} {\bibfnamefont {S.~A.}\ \bibnamefont
  {Sato}}, \bibinfo {author} {\bibfnamefont {J.~W.}\ \bibnamefont {McIver}},
  \bibinfo {author} {\bibfnamefont {M.}~\bibnamefont {Nuske}}, \bibinfo
  {author} {\bibfnamefont {P.}~\bibnamefont {Tang}}, \bibinfo {author}
  {\bibfnamefont {G.}~\bibnamefont {Jotzu}}, \bibinfo {author} {\bibfnamefont
  {B.}~\bibnamefont {Schulte}}, \bibinfo {author} {\bibfnamefont
  {H.}~\bibnamefont {H{\"u}bener}}, \bibinfo {author} {\bibfnamefont
  {U.}~\bibnamefont {De~Giovannini}}, \bibinfo {author} {\bibfnamefont
  {L.}~\bibnamefont {Mathey}}, \bibinfo {author} {\bibfnamefont {M.~A.}\
  \bibnamefont {Sentef}}, \emph {et~al.},\ }\bibfield  {title} {\bibinfo
  {title} {Microscopic theory for the light-induced anomalous {H}all effect in
  graphene},\ }\href@noop {} {\bibfield  {journal} {\bibinfo  {journal} {Phys.
  Rev. B}\ }\textbf {\bibinfo {volume} {99}},\ \bibinfo {pages} {214302}
  (\bibinfo {year} {2019})}\BibitemShut {NoStop}%
\bibitem [{\citenamefont {Earl}\ \emph {et~al.}(2021)\citenamefont {Earl},
  \citenamefont {Conway}, \citenamefont {Muir}, \citenamefont {Wurdack},
  \citenamefont {Ostrovskaya}, \citenamefont {Tollerud},\ and\ \citenamefont
  {Davis}}]{earl2021coherent}%
  \BibitemOpen
  \bibfield  {author} {\bibinfo {author} {\bibfnamefont {S.~K.}\ \bibnamefont
  {Earl}}, \bibinfo {author} {\bibfnamefont {M.~A.}\ \bibnamefont {Conway}},
  \bibinfo {author} {\bibfnamefont {J.~B.}\ \bibnamefont {Muir}}, \bibinfo
  {author} {\bibfnamefont {M.}~\bibnamefont {Wurdack}}, \bibinfo {author}
  {\bibfnamefont {E.~A.}\ \bibnamefont {Ostrovskaya}}, \bibinfo {author}
  {\bibfnamefont {J.~O.}\ \bibnamefont {Tollerud}},\ and\ \bibinfo {author}
  {\bibfnamefont {J.~A.}\ \bibnamefont {Davis}},\ }\bibfield  {title} {\bibinfo
  {title} {Coherent dynamics of {F}loquet-{B}loch states in monolayer {WS}$_2$
  reveals fast adiabatic switching},\ }\href@noop {} {\bibfield  {journal}
  {\bibinfo  {journal} {Phys. Rev. B}\ }\textbf {\bibinfo {volume} {104}},\
  \bibinfo {pages} {L060303} (\bibinfo {year} {2021})}\BibitemShut {NoStop}%
\bibitem [{\citenamefont {Michael}\ \emph
  {et~al.}(2022{\natexlab{a}})\citenamefont {Michael}, \citenamefont
  {F{\"o}rst}, \citenamefont {Nicoletti}, \citenamefont {Haque}, \citenamefont
  {Zhang}, \citenamefont {Cavalleri}, \citenamefont {Averitt}, \citenamefont
  {Podolsky},\ and\ \citenamefont {Demler}}]{michael2022generalized}%
  \BibitemOpen
  \bibfield  {author} {\bibinfo {author} {\bibfnamefont {M.~H.}\ \bibnamefont
  {Michael}}, \bibinfo {author} {\bibfnamefont {M.}~\bibnamefont {F{\"o}rst}},
  \bibinfo {author} {\bibfnamefont {D.}~\bibnamefont {Nicoletti}}, \bibinfo
  {author} {\bibfnamefont {S.~R.~U.}\ \bibnamefont {Haque}}, \bibinfo {author}
  {\bibfnamefont {Y.}~\bibnamefont {Zhang}}, \bibinfo {author} {\bibfnamefont
  {A.}~\bibnamefont {Cavalleri}}, \bibinfo {author} {\bibfnamefont {R.~D.}\
  \bibnamefont {Averitt}}, \bibinfo {author} {\bibfnamefont {D.}~\bibnamefont
  {Podolsky}},\ and\ \bibinfo {author} {\bibfnamefont {E.}~\bibnamefont
  {Demler}},\ }\bibfield  {title} {\bibinfo {title} {Generalized
  {F}resnel-{F}loquet equations for driven quantum materials},\ }\href@noop {}
  {\bibfield  {journal} {\bibinfo  {journal} {Phys. Rev. B}\ }\textbf {\bibinfo
  {volume} {105}},\ \bibinfo {pages} {174301} (\bibinfo {year}
  {2022}{\natexlab{a}})}\BibitemShut {NoStop}%
\bibitem [{\citenamefont {Michael}\ \emph
  {et~al.}(2022{\natexlab{b}})\citenamefont {Michael}, \citenamefont {Haque},
  \citenamefont {Windgaetter}, \citenamefont {Latini}, \citenamefont {Zhang},
  \citenamefont {Rubio}, \citenamefont {Averitt},\ and\ \citenamefont
  {Demler}}]{michael2022fresnel}%
  \BibitemOpen
  \bibfield  {author} {\bibinfo {author} {\bibfnamefont {M.}~\bibnamefont
  {Michael}}, \bibinfo {author} {\bibfnamefont {S.~R.~U.}\ \bibnamefont
  {Haque}}, \bibinfo {author} {\bibfnamefont {L.}~\bibnamefont {Windgaetter}},
  \bibinfo {author} {\bibfnamefont {S.}~\bibnamefont {Latini}}, \bibinfo
  {author} {\bibfnamefont {Y.}~\bibnamefont {Zhang}}, \bibinfo {author}
  {\bibfnamefont {A.}~\bibnamefont {Rubio}}, \bibinfo {author} {\bibfnamefont
  {R.~D.}\ \bibnamefont {Averitt}},\ and\ \bibinfo {author} {\bibfnamefont
  {E.}~\bibnamefont {Demler}},\ }\bibfield  {title} {\bibinfo {title}
  {Fresnel-{F}loquet theory of light-induced terahertz reflectivity
  amplification in {T}a$_2${N}i{S}e$_5$},\ }\href@noop {} {\bibfield  {journal}
  {\bibinfo  {journal} {arXiv:2207.08851}\ } (\bibinfo {year}
  {2022}{\natexlab{b}})}\BibitemShut {NoStop}%
\bibitem [{\citenamefont {Dehghani}\ and\ \citenamefont
  {Mitra}(2015)}]{dehghani2015optical}%
  \BibitemOpen
  \bibfield  {author} {\bibinfo {author} {\bibfnamefont {H.}~\bibnamefont
  {Dehghani}}\ and\ \bibinfo {author} {\bibfnamefont {A.}~\bibnamefont
  {Mitra}},\ }\bibfield  {title} {\bibinfo {title} {Optical {H}all conductivity
  of a {F}loquet topological insulator},\ }\href@noop {} {\bibfield  {journal}
  {\bibinfo  {journal} {Phys. Rev. B}\ }\textbf {\bibinfo {volume} {92}},\
  \bibinfo {pages} {165111} (\bibinfo {year} {2015})}\BibitemShut {NoStop}%
\bibitem [{\citenamefont {Du}\ \emph {et~al.}(2017)\citenamefont {Du},
  \citenamefont {Zhou},\ and\ \citenamefont {Fiete}}]{du2017quadratic}%
  \BibitemOpen
  \bibfield  {author} {\bibinfo {author} {\bibfnamefont {L.}~\bibnamefont
  {Du}}, \bibinfo {author} {\bibfnamefont {X.}~\bibnamefont {Zhou}},\ and\
  \bibinfo {author} {\bibfnamefont {G.~A.}\ \bibnamefont {Fiete}},\ }\bibfield
  {title} {\bibinfo {title} {Quadratic band touching points and flat bands in
  two-dimensional topological {F}loquet systems},\ }\href@noop {} {\bibfield
  {journal} {\bibinfo  {journal} {Phys. Rev. B}\ }\textbf {\bibinfo {volume}
  {95}},\ \bibinfo {pages} {035136} (\bibinfo {year} {2017})}\BibitemShut
  {NoStop}%
\bibitem [{\citenamefont {Chen}\ \emph {et~al.}(2018)\citenamefont {Chen},
  \citenamefont {Du},\ and\ \citenamefont {Fiete}}]{chen2018floquet}%
  \BibitemOpen
  \bibfield  {author} {\bibinfo {author} {\bibfnamefont {Q.}~\bibnamefont
  {Chen}}, \bibinfo {author} {\bibfnamefont {L.}~\bibnamefont {Du}},\ and\
  \bibinfo {author} {\bibfnamefont {G.~A.}\ \bibnamefont {Fiete}},\ }\bibfield
  {title} {\bibinfo {title} {Floquet band structure of a semi-{D}irac system},\
  }\href@noop {} {\bibfield  {journal} {\bibinfo  {journal} {Phys. Rev. B}\
  }\textbf {\bibinfo {volume} {97}},\ \bibinfo {pages} {035422} (\bibinfo
  {year} {2018})}\BibitemShut {NoStop}%
\bibitem [{\citenamefont {Kumar}\ \emph {et~al.}(2020)\citenamefont {Kumar},
  \citenamefont {Rodriguez-Vega}, \citenamefont {Pereg-Barnea},\ and\
  \citenamefont {Seradjeh}}]{kumar2020linear}%
  \BibitemOpen
  \bibfield  {author} {\bibinfo {author} {\bibfnamefont {A.}~\bibnamefont
  {Kumar}}, \bibinfo {author} {\bibfnamefont {M.}~\bibnamefont
  {Rodriguez-Vega}}, \bibinfo {author} {\bibfnamefont {T.}~\bibnamefont
  {Pereg-Barnea}},\ and\ \bibinfo {author} {\bibfnamefont {B.}~\bibnamefont
  {Seradjeh}},\ }\bibfield  {title} {\bibinfo {title} {Linear response theory
  and optical conductivity of {F}loquet topological insulators},\ }\href@noop
  {} {\bibfield  {journal} {\bibinfo  {journal} {Phys. Rev. B}\ }\textbf
  {\bibinfo {volume} {101}},\ \bibinfo {pages} {174314} (\bibinfo {year}
  {2020})}\BibitemShut {NoStop}%
\bibitem [{\citenamefont {Broers}\ and\ \citenamefont
  {Mathey}(2021)}]{broers2021observing}%
  \BibitemOpen
  \bibfield  {author} {\bibinfo {author} {\bibfnamefont {L.}~\bibnamefont
  {Broers}}\ and\ \bibinfo {author} {\bibfnamefont {L.}~\bibnamefont
  {Mathey}},\ }\bibfield  {title} {\bibinfo {title} {Observing light-induced
  {F}loquet band gaps in the longitudinal conductivity of graphene},\
  }\href@noop {} {\bibfield  {journal} {\bibinfo  {journal} {Commun. Phys.}\
  }\textbf {\bibinfo {volume} {4}},\ \bibinfo {pages} {1} (\bibinfo {year}
  {2021})}\BibitemShut {NoStop}%
\bibitem [{\citenamefont {Eckhardt}\ \emph {et~al.}(2022)\citenamefont
  {Eckhardt}, \citenamefont {Passetti}, \citenamefont {Othman}, \citenamefont
  {Karrasch}, \citenamefont {Cavaliere}, \citenamefont {Sentef},\ and\
  \citenamefont {Kennes}}]{eckhardt2022quantum}%
  \BibitemOpen
  \bibfield  {author} {\bibinfo {author} {\bibfnamefont {C.~J.}\ \bibnamefont
  {Eckhardt}}, \bibinfo {author} {\bibfnamefont {G.}~\bibnamefont {Passetti}},
  \bibinfo {author} {\bibfnamefont {M.}~\bibnamefont {Othman}}, \bibinfo
  {author} {\bibfnamefont {C.}~\bibnamefont {Karrasch}}, \bibinfo {author}
  {\bibfnamefont {F.}~\bibnamefont {Cavaliere}}, \bibinfo {author}
  {\bibfnamefont {M.~A.}\ \bibnamefont {Sentef}},\ and\ \bibinfo {author}
  {\bibfnamefont {D.~M.}\ \bibnamefont {Kennes}},\ }\bibfield  {title}
  {\bibinfo {title} {Quantum {F}loquet engineering with an exactly solvable
  tight-binding chain in a cavity},\ }\href@noop {} {\bibfield  {journal}
  {\bibinfo  {journal} {Commun. Phys.}\ }\textbf {\bibinfo {volume} {5}},\
  \bibinfo {pages} {1} (\bibinfo {year} {2022})}\BibitemShut {NoStop}%
\bibitem [{\citenamefont {Ahmadabadi}\ \emph {et~al.}(2022)\citenamefont
  {Ahmadabadi}, \citenamefont {Dehghani},\ and\ \citenamefont
  {Hafezi}}]{ahmadabadi2022optical}%
  \BibitemOpen
  \bibfield  {author} {\bibinfo {author} {\bibfnamefont {I.}~\bibnamefont
  {Ahmadabadi}}, \bibinfo {author} {\bibfnamefont {H.}~\bibnamefont
  {Dehghani}},\ and\ \bibinfo {author} {\bibfnamefont {M.}~\bibnamefont
  {Hafezi}},\ }\bibfield  {title} {\bibinfo {title} {Optical conductivity and
  orbital magnetization of {F}loquet vortex states},\ }\href@noop {} {\bibfield
   {journal} {\bibinfo  {journal} {arXiv:2204.09488}\ } (\bibinfo {year}
  {2022})}\BibitemShut {NoStop}%
\bibitem [{\citenamefont {Dabiri}\ \emph {et~al.}(2022)\citenamefont {Dabiri},
  \citenamefont {Cheraghchi},\ and\ \citenamefont
  {Sadeghi}}]{dabiri2022floquet}%
  \BibitemOpen
  \bibfield  {author} {\bibinfo {author} {\bibfnamefont {S.~S.}\ \bibnamefont
  {Dabiri}}, \bibinfo {author} {\bibfnamefont {H.}~\bibnamefont {Cheraghchi}},\
  and\ \bibinfo {author} {\bibfnamefont {A.}~\bibnamefont {Sadeghi}},\
  }\bibfield  {title} {\bibinfo {title} {Floquet states and optical
  conductivity of an irradiated two-dimensional topological insulator},\
  }\href@noop {} {\bibfield  {journal} {\bibinfo  {journal} {Phys. Rev. B}\
  }\textbf {\bibinfo {volume} {106}},\ \bibinfo {pages} {165423} (\bibinfo
  {year} {2022})}\BibitemShut {NoStop}%
\bibitem [{\citenamefont {Abergel}\ \emph {et~al.}(2007)\citenamefont
  {Abergel}, \citenamefont {Russell},\ and\ \citenamefont
  {Fal’ko}}]{abergel2007visibility}%
  \BibitemOpen
  \bibfield  {author} {\bibinfo {author} {\bibfnamefont {D.~S.~L.}\
  \bibnamefont {Abergel}}, \bibinfo {author} {\bibfnamefont {A.}~\bibnamefont
  {Russell}},\ and\ \bibinfo {author} {\bibfnamefont {V.~I.}\ \bibnamefont
  {Fal’ko}},\ }\bibfield  {title} {\bibinfo {title} {Visibility of graphene
  flakes on a dielectric substrate},\ }\href@noop {} {\bibfield  {journal}
  {\bibinfo  {journal} {Appl. Phys. Lett.}\ }\textbf {\bibinfo {volume} {91}},\
  \bibinfo {pages} {063125} (\bibinfo {year} {2007})}\BibitemShut {NoStop}%
\bibitem [{\citenamefont {Yoo}\ and\ \citenamefont
  {Park}(2022)}]{yoo2022spectroscopic}%
  \BibitemOpen
  \bibfield  {author} {\bibinfo {author} {\bibfnamefont {S.}~\bibnamefont
  {Yoo}}\ and\ \bibinfo {author} {\bibfnamefont {Q.-H.}\ \bibnamefont {Park}},\
  }\bibfield  {title} {\bibinfo {title} {Spectroscopic ellipsometry for
  low-dimensional materials and heterostructures},\ }\href@noop {} {\bibfield
  {journal} {\bibinfo  {journal} {Nanophotonics}\ }\textbf {\bibinfo {volume}
  {11}},\ \bibinfo {pages} {2811} (\bibinfo {year} {2022})}\BibitemShut
  {NoStop}%
\bibitem [{\citenamefont {Roberts}\ \emph {et~al.}(2011)\citenamefont
  {Roberts}, \citenamefont {Cormode}, \citenamefont {Reynolds}, \citenamefont
  {Newhouse-Illige}, \citenamefont {LeRoy},\ and\ \citenamefont
  {Sandhu}}]{roberts2011response}%
  \BibitemOpen
  \bibfield  {author} {\bibinfo {author} {\bibfnamefont {A.}~\bibnamefont
  {Roberts}}, \bibinfo {author} {\bibfnamefont {D.}~\bibnamefont {Cormode}},
  \bibinfo {author} {\bibfnamefont {C.}~\bibnamefont {Reynolds}}, \bibinfo
  {author} {\bibfnamefont {T.}~\bibnamefont {Newhouse-Illige}}, \bibinfo
  {author} {\bibfnamefont {B.~J.}\ \bibnamefont {LeRoy}},\ and\ \bibinfo
  {author} {\bibfnamefont {A.~S.}\ \bibnamefont {Sandhu}},\ }\bibfield  {title}
  {\bibinfo {title} {Response of graphene to femtosecond high-intensity laser
  irradiation},\ }\href@noop {} {\bibfield  {journal} {\bibinfo  {journal}
  {Appl. Phys. Lett.}\ }\textbf {\bibinfo {volume} {99}},\ \bibinfo {pages}
  {051912} (\bibinfo {year} {2011})}\BibitemShut {NoStop}%
\bibitem [{\citenamefont {Cupo}\ \emph {et~al.}(2021)\citenamefont {Cupo},
  \citenamefont {Cobanera}, \citenamefont {Whitfield}, \citenamefont
  {Ramanathan},\ and\ \citenamefont {Viola}}]{cupo2021floquet}%
  \BibitemOpen
  \bibfield  {author} {\bibinfo {author} {\bibfnamefont {A.}~\bibnamefont
  {Cupo}}, \bibinfo {author} {\bibfnamefont {E.}~\bibnamefont {Cobanera}},
  \bibinfo {author} {\bibfnamefont {J.~D.}\ \bibnamefont {Whitfield}}, \bibinfo
  {author} {\bibfnamefont {C.}~\bibnamefont {Ramanathan}},\ and\ \bibinfo
  {author} {\bibfnamefont {L.}~\bibnamefont {Viola}},\ }\bibfield  {title}
  {\bibinfo {title} {Floquet graphene antidot lattices},\ }\href@noop {}
  {\bibfield  {journal} {\bibinfo  {journal} {Phys. Rev. B}\ }\textbf {\bibinfo
  {volume} {104}},\ \bibinfo {pages} {174304} (\bibinfo {year}
  {2021})}\BibitemShut {NoStop}%
\bibitem [{\citenamefont {Bai}\ \emph {et~al.}(2010)\citenamefont {Bai},
  \citenamefont {Zhong}, \citenamefont {Jiang}, \citenamefont {Huang},\ and\
  \citenamefont {Duan}}]{bai2010graphene}%
  \BibitemOpen
  \bibfield  {author} {\bibinfo {author} {\bibfnamefont {J.}~\bibnamefont
  {Bai}}, \bibinfo {author} {\bibfnamefont {X.}~\bibnamefont {Zhong}}, \bibinfo
  {author} {\bibfnamefont {S.}~\bibnamefont {Jiang}}, \bibinfo {author}
  {\bibfnamefont {Y.}~\bibnamefont {Huang}},\ and\ \bibinfo {author}
  {\bibfnamefont {X.}~\bibnamefont {Duan}},\ }\bibfield  {title} {\bibinfo
  {title} {Graphene nanomesh},\ }\href@noop {} {\bibfield  {journal} {\bibinfo
  {journal} {Nat. Nanotechnol.}\ }\textbf {\bibinfo {volume} {5}},\ \bibinfo
  {pages} {190} (\bibinfo {year} {2010})}\BibitemShut {NoStop}%
\bibitem [{\citenamefont {Jessen}\ \emph {et~al.}(2019)\citenamefont {Jessen},
  \citenamefont {Gammelgaard}, \citenamefont {Thomsen}, \citenamefont
  {Mackenzie}, \citenamefont {Thomsen}, \citenamefont {Caridad}, \citenamefont
  {Duegaard}, \citenamefont {Watanabe}, \citenamefont {Taniguchi},
  \citenamefont {Booth} \emph {et~al.}}]{jessen2019lithographic}%
  \BibitemOpen
  \bibfield  {author} {\bibinfo {author} {\bibfnamefont {B.~S.}\ \bibnamefont
  {Jessen}}, \bibinfo {author} {\bibfnamefont {L.}~\bibnamefont {Gammelgaard}},
  \bibinfo {author} {\bibfnamefont {M.~R.}\ \bibnamefont {Thomsen}}, \bibinfo
  {author} {\bibfnamefont {D.~M.~A.}\ \bibnamefont {Mackenzie}}, \bibinfo
  {author} {\bibfnamefont {J.~D.}\ \bibnamefont {Thomsen}}, \bibinfo {author}
  {\bibfnamefont {J.~M.}\ \bibnamefont {Caridad}}, \bibinfo {author}
  {\bibfnamefont {E.}~\bibnamefont {Duegaard}}, \bibinfo {author}
  {\bibfnamefont {K.}~\bibnamefont {Watanabe}}, \bibinfo {author}
  {\bibfnamefont {T.}~\bibnamefont {Taniguchi}}, \bibinfo {author}
  {\bibfnamefont {T.~J.}\ \bibnamefont {Booth}}, \emph {et~al.},\ }\bibfield
  {title} {\bibinfo {title} {Lithographic band structure engineering of
  graphene},\ }\href@noop {} {\bibfield  {journal} {\bibinfo  {journal} {Nat.
  Nanotechnol.}\ }\textbf {\bibinfo {volume} {14}},\ \bibinfo {pages} {340}
  (\bibinfo {year} {2019})}\BibitemShut {NoStop}%
\bibitem [{\citenamefont {Moreno}\ \emph {et~al.}(2018)\citenamefont {Moreno},
  \citenamefont {Vilas-Varela}, \citenamefont {Kretz}, \citenamefont
  {Garcia-Lekue}, \citenamefont {Costache}, \citenamefont {Paradinas},
  \citenamefont {Panighel}, \citenamefont {Ceballos}, \citenamefont
  {Valenzuela}, \citenamefont {Pe{\~n}a} \emph {et~al.}}]{moreno2018bottom}%
  \BibitemOpen
  \bibfield  {author} {\bibinfo {author} {\bibfnamefont {C.}~\bibnamefont
  {Moreno}}, \bibinfo {author} {\bibfnamefont {M.}~\bibnamefont
  {Vilas-Varela}}, \bibinfo {author} {\bibfnamefont {B.}~\bibnamefont {Kretz}},
  \bibinfo {author} {\bibfnamefont {A.}~\bibnamefont {Garcia-Lekue}}, \bibinfo
  {author} {\bibfnamefont {M.~V.}\ \bibnamefont {Costache}}, \bibinfo {author}
  {\bibfnamefont {M.}~\bibnamefont {Paradinas}}, \bibinfo {author}
  {\bibfnamefont {M.}~\bibnamefont {Panighel}}, \bibinfo {author}
  {\bibfnamefont {G.}~\bibnamefont {Ceballos}}, \bibinfo {author}
  {\bibfnamefont {S.~O.}\ \bibnamefont {Valenzuela}}, \bibinfo {author}
  {\bibfnamefont {D.}~\bibnamefont {Pe{\~n}a}}, \emph {et~al.},\ }\bibfield
  {title} {\bibinfo {title} {Bottom-up synthesis of multifunctional nanoporous
  graphene},\ }\href@noop {} {\bibfield  {journal} {\bibinfo  {journal}
  {Science}\ }\textbf {\bibinfo {volume} {360}},\ \bibinfo {pages} {199}
  (\bibinfo {year} {2018})}\BibitemShut {NoStop}%
\bibitem [{\citenamefont {Pedersen}\ \emph
  {et~al.}(2008{\natexlab{a}})\citenamefont {Pedersen}, \citenamefont {Flindt},
  \citenamefont {Pedersen}, \citenamefont {Mortensen}, \citenamefont {Jauho},\
  and\ \citenamefont {Pedersen}}]{pedersen2008graphene}%
  \BibitemOpen
  \bibfield  {author} {\bibinfo {author} {\bibfnamefont {T.~G.}\ \bibnamefont
  {Pedersen}}, \bibinfo {author} {\bibfnamefont {C.}~\bibnamefont {Flindt}},
  \bibinfo {author} {\bibfnamefont {J.}~\bibnamefont {Pedersen}}, \bibinfo
  {author} {\bibfnamefont {N.~A.}\ \bibnamefont {Mortensen}}, \bibinfo {author}
  {\bibfnamefont {A.-P.}\ \bibnamefont {Jauho}},\ and\ \bibinfo {author}
  {\bibfnamefont {K.}~\bibnamefont {Pedersen}},\ }\bibfield  {title} {\bibinfo
  {title} {Graphene antidot lattices: {D}esigned defects and spin qubits},\
  }\href@noop {} {\bibfield  {journal} {\bibinfo  {journal} {Phys. Rev. Lett.}\
  }\textbf {\bibinfo {volume} {100}},\ \bibinfo {pages} {136804} (\bibinfo
  {year} {2008}{\natexlab{a}})}\BibitemShut {NoStop}%
\bibitem [{\citenamefont {F{\"u}rst}\ \emph {et~al.}(2009)\citenamefont
  {F{\"u}rst}, \citenamefont {Pedersen}, \citenamefont {Flindt}, \citenamefont
  {Mortensen}, \citenamefont {Brandbyge}, \citenamefont {Pedersen},\ and\
  \citenamefont {Jauho}}]{furst2009electronic}%
  \BibitemOpen
  \bibfield  {author} {\bibinfo {author} {\bibfnamefont {J.~A.}\ \bibnamefont
  {F{\"u}rst}}, \bibinfo {author} {\bibfnamefont {J.~G.}\ \bibnamefont
  {Pedersen}}, \bibinfo {author} {\bibfnamefont {C.}~\bibnamefont {Flindt}},
  \bibinfo {author} {\bibfnamefont {N.~A.}\ \bibnamefont {Mortensen}}, \bibinfo
  {author} {\bibfnamefont {M.}~\bibnamefont {Brandbyge}}, \bibinfo {author}
  {\bibfnamefont {T.~G.}\ \bibnamefont {Pedersen}},\ and\ \bibinfo {author}
  {\bibfnamefont {A.-P.}\ \bibnamefont {Jauho}},\ }\bibfield  {title} {\bibinfo
  {title} {Electronic properties of graphene antidot lattices},\ }\href@noop {}
  {\bibfield  {journal} {\bibinfo  {journal} {New J. Phys.}\ }\textbf {\bibinfo
  {volume} {11}},\ \bibinfo {pages} {095020} (\bibinfo {year}
  {2009})}\BibitemShut {NoStop}%
\bibitem [{\citenamefont {Brun}\ \emph {et~al.}(2014)\citenamefont {Brun},
  \citenamefont {Thomsen},\ and\ \citenamefont
  {Pedersen}}]{brun2014electronic}%
  \BibitemOpen
  \bibfield  {author} {\bibinfo {author} {\bibfnamefont {S.~J.}\ \bibnamefont
  {Brun}}, \bibinfo {author} {\bibfnamefont {M.~R.}\ \bibnamefont {Thomsen}},\
  and\ \bibinfo {author} {\bibfnamefont {T.~G.}\ \bibnamefont {Pedersen}},\
  }\bibfield  {title} {\bibinfo {title} {Electronic and optical properties of
  graphene antidot lattices: {C}omparison of {D}irac and tight-binding
  models},\ }\href@noop {} {\bibfield  {journal} {\bibinfo  {journal} {J. Phys.
  Condens. Matter}\ }\textbf {\bibinfo {volume} {26}},\ \bibinfo {pages}
  {265301} (\bibinfo {year} {2014})}\BibitemShut {NoStop}%
\bibitem [{\citenamefont {Gmitra}\ \emph {et~al.}(2009)\citenamefont {Gmitra},
  \citenamefont {Konschuh}, \citenamefont {Ertler}, \citenamefont
  {Ambrosch-Draxl},\ and\ \citenamefont {Fabian}}]{gmitra2009band}%
  \BibitemOpen
  \bibfield  {author} {\bibinfo {author} {\bibfnamefont {M.}~\bibnamefont
  {Gmitra}}, \bibinfo {author} {\bibfnamefont {S.}~\bibnamefont {Konschuh}},
  \bibinfo {author} {\bibfnamefont {C.}~\bibnamefont {Ertler}}, \bibinfo
  {author} {\bibfnamefont {C.}~\bibnamefont {Ambrosch-Draxl}},\ and\ \bibinfo
  {author} {\bibfnamefont {J.}~\bibnamefont {Fabian}},\ }\bibfield  {title}
  {\bibinfo {title} {Band-structure topologies of graphene: Spin-orbit coupling
  effects from first principles},\ }\href@noop {} {\bibfield  {journal}
  {\bibinfo  {journal} {Phys. Rev. B}\ }\textbf {\bibinfo {volume} {80}},\
  \bibinfo {pages} {235431} (\bibinfo {year} {2009})}\BibitemShut {NoStop}%
\bibitem [{\citenamefont {Konschuh}\ \emph {et~al.}(2010)\citenamefont
  {Konschuh}, \citenamefont {Gmitra},\ and\ \citenamefont
  {Fabian}}]{konschuh2010tight}%
  \BibitemOpen
  \bibfield  {author} {\bibinfo {author} {\bibfnamefont {S.}~\bibnamefont
  {Konschuh}}, \bibinfo {author} {\bibfnamefont {M.}~\bibnamefont {Gmitra}},\
  and\ \bibinfo {author} {\bibfnamefont {J.}~\bibnamefont {Fabian}},\
  }\bibfield  {title} {\bibinfo {title} {Tight-binding theory of the spin-orbit
  coupling in graphene},\ }\href@noop {} {\bibfield  {journal} {\bibinfo
  {journal} {Phys. Rev. B}\ }\textbf {\bibinfo {volume} {82}},\ \bibinfo
  {pages} {245412} (\bibinfo {year} {2010})}\BibitemShut {NoStop}%
\bibitem [{\citenamefont {Li}\ \emph {et~al.}(2020)\citenamefont {Li},
  \citenamefont {Golez}, \citenamefont {Mazza}, \citenamefont {Millis},
  \citenamefont {Georges},\ and\ \citenamefont
  {Eckstein}}]{li2020electromagnetic}%
  \BibitemOpen
  \bibfield  {author} {\bibinfo {author} {\bibfnamefont {J.}~\bibnamefont
  {Li}}, \bibinfo {author} {\bibfnamefont {D.}~\bibnamefont {Golez}}, \bibinfo
  {author} {\bibfnamefont {G.}~\bibnamefont {Mazza}}, \bibinfo {author}
  {\bibfnamefont {A.~J.}\ \bibnamefont {Millis}}, \bibinfo {author}
  {\bibfnamefont {A.}~\bibnamefont {Georges}},\ and\ \bibinfo {author}
  {\bibfnamefont {M.}~\bibnamefont {Eckstein}},\ }\bibfield  {title} {\bibinfo
  {title} {Electromagnetic coupling in tight-binding models for strongly
  correlated light and matter},\ }\href@noop {} {\bibfield  {journal} {\bibinfo
   {journal} {Phys. Rev. B}\ }\textbf {\bibinfo {volume} {101}},\ \bibinfo
  {pages} {205140} (\bibinfo {year} {2020})}\BibitemShut {NoStop}%
\bibitem [{\citenamefont {Vogl}\ \emph {et~al.}(2020)\citenamefont {Vogl},
  \citenamefont {Rodriguez-Vega},\ and\ \citenamefont
  {Fiete}}]{vogl2020unfold}%
  \BibitemOpen
  \bibfield  {author} {\bibinfo {author} {\bibfnamefont {M.}~\bibnamefont
  {Vogl}}, \bibinfo {author} {\bibfnamefont {M.}~\bibnamefont
  {Rodriguez-Vega}},\ and\ \bibinfo {author} {\bibfnamefont {G.~A.}\
  \bibnamefont {Fiete}},\ }\bibfield  {title} {\bibinfo {title} {Effective
  {F}loquet {H}amiltonian in the low-frequency regime},\ }\href@noop {}
  {\bibfield  {journal} {\bibinfo  {journal} {Phys. Rev. B}\ }\textbf {\bibinfo
  {volume} {101}},\ \bibinfo {pages} {024303} (\bibinfo {year}
  {2020})}\BibitemShut {NoStop}%
\bibitem [{\citenamefont {Torres}\ and\ \citenamefont
  {Kunold}(2005)}]{torres2005kubo}%
  \BibitemOpen
  \bibfield  {author} {\bibinfo {author} {\bibfnamefont {M.}~\bibnamefont
  {Torres}}\ and\ \bibinfo {author} {\bibfnamefont {A.}~\bibnamefont
  {Kunold}},\ }\bibfield  {title} {\bibinfo {title} {Kubo formula for {F}loquet
  states and photoconductivity oscillations in a two-dimensional electron
  gas},\ }\href@noop {} {\bibfield  {journal} {\bibinfo  {journal} {Phys. Rev.
  B}\ }\textbf {\bibinfo {volume} {71}},\ \bibinfo {pages} {115313} (\bibinfo
  {year} {2005})}\BibitemShut {NoStop}%
\bibitem [{\citenamefont {Oka}\ and\ \citenamefont
  {Aoki}(2009)}]{oka2009photovoltaic}%
  \BibitemOpen
  \bibfield  {author} {\bibinfo {author} {\bibfnamefont {T.}~\bibnamefont
  {Oka}}\ and\ \bibinfo {author} {\bibfnamefont {H.}~\bibnamefont {Aoki}},\
  }\bibfield  {title} {\bibinfo {title} {Photovoltaic {H}all effect in
  graphene},\ }\href@noop {} {\bibfield  {journal} {\bibinfo  {journal} {Phys.
  Rev. B}\ }\textbf {\bibinfo {volume} {79}},\ \bibinfo {pages} {081406(R)}
  (\bibinfo {year} {2009})}\BibitemShut {NoStop}%
\bibitem [{\citenamefont {Dehghani}\ \emph {et~al.}(2015)\citenamefont
  {Dehghani}, \citenamefont {Oka},\ and\ \citenamefont
  {Mitra}}]{dehghani2015out}%
  \BibitemOpen
  \bibfield  {author} {\bibinfo {author} {\bibfnamefont {H.}~\bibnamefont
  {Dehghani}}, \bibinfo {author} {\bibfnamefont {T.}~\bibnamefont {Oka}},\ and\
  \bibinfo {author} {\bibfnamefont {A.}~\bibnamefont {Mitra}},\ }\bibfield
  {title} {\bibinfo {title} {Out-of-equilibrium electrons and the {H}all
  conductance of a {F}loquet topological insulator},\ }\href@noop {} {\bibfield
   {journal} {\bibinfo  {journal} {Phys. Rev. B}\ }\textbf {\bibinfo {volume}
  {91}},\ \bibinfo {pages} {155422} (\bibinfo {year} {2015})}\BibitemShut
  {NoStop}%
\bibitem [{\citenamefont {Oka}\ and\ \citenamefont
  {Bucciantini}(2016)}]{oka2016heterodyne}%
  \BibitemOpen
  \bibfield  {author} {\bibinfo {author} {\bibfnamefont {T.}~\bibnamefont
  {Oka}}\ and\ \bibinfo {author} {\bibfnamefont {L.}~\bibnamefont
  {Bucciantini}},\ }\bibfield  {title} {\bibinfo {title} {Heterodyne {H}all
  effect in a two-dimensional electron gas},\ }\href@noop {} {\bibfield
  {journal} {\bibinfo  {journal} {Phys. Rev. B}\ }\textbf {\bibinfo {volume}
  {94}},\ \bibinfo {pages} {155133} (\bibinfo {year} {2016})}\BibitemShut
  {NoStop}%
\bibitem [{\citenamefont {Wackerl}\ \emph {et~al.}(2020)\citenamefont
  {Wackerl}, \citenamefont {Wenk},\ and\ \citenamefont
  {Schliemann}}]{wackerl2020floquet}%
  \BibitemOpen
  \bibfield  {author} {\bibinfo {author} {\bibfnamefont {M.}~\bibnamefont
  {Wackerl}}, \bibinfo {author} {\bibfnamefont {P.}~\bibnamefont {Wenk}},\ and\
  \bibinfo {author} {\bibfnamefont {J.}~\bibnamefont {Schliemann}},\ }\bibfield
   {title} {\bibinfo {title} {Floquet-{D}rude conductivity},\ }\href@noop {}
  {\bibfield  {journal} {\bibinfo  {journal} {Phys. Rev. B}\ }\textbf {\bibinfo
  {volume} {101}},\ \bibinfo {pages} {184204} (\bibinfo {year}
  {2020})}\BibitemShut {NoStop}%
\bibitem [{\citenamefont {H{\"u}bener}\ \emph {et~al.}(2017)\citenamefont
  {H{\"u}bener}, \citenamefont {Sentef}, \citenamefont {De~Giovannini},
  \citenamefont {Kemper},\ and\ \citenamefont {Rubio}}]{hubener2017creating}%
  \BibitemOpen
  \bibfield  {author} {\bibinfo {author} {\bibfnamefont {H.}~\bibnamefont
  {H{\"u}bener}}, \bibinfo {author} {\bibfnamefont {M.~A.}\ \bibnamefont
  {Sentef}}, \bibinfo {author} {\bibfnamefont {U.}~\bibnamefont
  {De~Giovannini}}, \bibinfo {author} {\bibfnamefont {A.~F.}\ \bibnamefont
  {Kemper}},\ and\ \bibinfo {author} {\bibfnamefont {A.}~\bibnamefont
  {Rubio}},\ }\bibfield  {title} {\bibinfo {title} {Creating stable
  {F}loquet--{W}eyl semimetals by laser-driving of 3{D} {D}irac materials},\
  }\href@noop {} {\bibfield  {journal} {\bibinfo  {journal} {Nat. Commun.}\
  }\textbf {\bibinfo {volume} {8}},\ \bibinfo {pages} {1} (\bibinfo {year}
  {2017})}\BibitemShut {NoStop}%
\bibitem [{\citenamefont {Park}\ \emph {et~al.}(2009)\citenamefont {Park},
  \citenamefont {Giustino}, \citenamefont {Spataru}, \citenamefont {Cohen},\
  and\ \citenamefont {Louie}}]{park2009first}%
  \BibitemOpen
  \bibfield  {author} {\bibinfo {author} {\bibfnamefont {C.-H.}\ \bibnamefont
  {Park}}, \bibinfo {author} {\bibfnamefont {F.}~\bibnamefont {Giustino}},
  \bibinfo {author} {\bibfnamefont {C.~D.}\ \bibnamefont {Spataru}}, \bibinfo
  {author} {\bibfnamefont {M.~L.}\ \bibnamefont {Cohen}},\ and\ \bibinfo
  {author} {\bibfnamefont {S.~G.}\ \bibnamefont {Louie}},\ }\bibfield  {title}
  {\bibinfo {title} {First-principles study of electron linewidths in
  graphene},\ }\href@noop {} {\bibfield  {journal} {\bibinfo  {journal} {Phys.
  Rev. Lett.}\ }\textbf {\bibinfo {volume} {102}},\ \bibinfo {pages} {076803}
  (\bibinfo {year} {2009})}\BibitemShut {NoStop}%
\bibitem [{\citenamefont {D’Alessio}\ and\ \citenamefont
  {Rigol}(2015)}]{d2015dynamical}%
  \BibitemOpen
  \bibfield  {author} {\bibinfo {author} {\bibfnamefont {L.}~\bibnamefont
  {D’Alessio}}\ and\ \bibinfo {author} {\bibfnamefont {M.}~\bibnamefont
  {Rigol}},\ }\bibfield  {title} {\bibinfo {title} {Dynamical preparation of
  {F}loquet {C}hern insulators},\ }\href@noop {} {\bibfield  {journal}
  {\bibinfo  {journal} {Nat. Commun.}\ }\textbf {\bibinfo {volume} {6}},\
  \bibinfo {pages} {1} (\bibinfo {year} {2015})}\BibitemShut {NoStop}%
\bibitem [{\citenamefont {Delplace}\ \emph {et~al.}(2013)\citenamefont
  {Delplace}, \citenamefont {G{\'o}mez-Le{\'o}n},\ and\ \citenamefont
  {Platero}}]{delplace2013merging}%
  \BibitemOpen
  \bibfield  {author} {\bibinfo {author} {\bibfnamefont {P.}~\bibnamefont
  {Delplace}}, \bibinfo {author} {\bibfnamefont {{\'A}.}~\bibnamefont
  {G{\'o}mez-Le{\'o}n}},\ and\ \bibinfo {author} {\bibfnamefont
  {G.}~\bibnamefont {Platero}},\ }\bibfield  {title} {\bibinfo {title} {Merging
  of {D}irac points and {F}loquet topological transitions in {AC}-driven
  graphene},\ }\href@noop {} {\bibfield  {journal} {\bibinfo  {journal} {Phys.
  Rev. B}\ }\textbf {\bibinfo {volume} {88}},\ \bibinfo {pages} {245422}
  (\bibinfo {year} {2013})}\BibitemShut {NoStop}%
\bibitem [{\citenamefont {Agarwala}\ \emph {et~al.}(2016)\citenamefont
  {Agarwala}, \citenamefont {Bhattacharya}, \citenamefont {Dutta},\ and\
  \citenamefont {Sen}}]{agarwala2016effects}%
  \BibitemOpen
  \bibfield  {author} {\bibinfo {author} {\bibfnamefont {A.}~\bibnamefont
  {Agarwala}}, \bibinfo {author} {\bibfnamefont {U.}~\bibnamefont
  {Bhattacharya}}, \bibinfo {author} {\bibfnamefont {A.}~\bibnamefont
  {Dutta}},\ and\ \bibinfo {author} {\bibfnamefont {D.}~\bibnamefont {Sen}},\
  }\bibfield  {title} {\bibinfo {title} {Effects of periodic kicking on
  dispersion and wave packet dynamics in graphene},\ }\href@noop {} {\bibfield
  {journal} {\bibinfo  {journal} {Phys. Rev. B}\ }\textbf {\bibinfo {volume}
  {93}},\ \bibinfo {pages} {174301} (\bibinfo {year} {2016})}\BibitemShut
  {NoStop}%
\bibitem [{\citenamefont {Liu}\ \emph {et~al.}(2019)\citenamefont {Liu},
  \citenamefont {Sun},\ and\ \citenamefont {Meng}}]{liu2019engineering}%
  \BibitemOpen
  \bibfield  {author} {\bibinfo {author} {\bibfnamefont {H.}~\bibnamefont
  {Liu}}, \bibinfo {author} {\bibfnamefont {J.-T.}\ \bibnamefont {Sun}},\ and\
  \bibinfo {author} {\bibfnamefont {S.}~\bibnamefont {Meng}},\ }\bibfield
  {title} {\bibinfo {title} {Engineering {D}irac states in graphene:
  {C}oexisting type-{I} and type-{II} {F}loquet-{D}irac fermions},\ }\href@noop
  {} {\bibfield  {journal} {\bibinfo  {journal} {Phys. Rev. B}\ }\textbf
  {\bibinfo {volume} {99}},\ \bibinfo {pages} {075121} (\bibinfo {year}
  {2019})}\BibitemShut {NoStop}%
\bibitem [{Note1()}]{Note1}%
  \BibitemOpen
  \bibinfo {note} {Note that all plotted conductivities in this paper are
  normalized to the number of spin channels per band (two)}\BibitemShut
  {NoStop}%
\bibitem [{\citenamefont {Pedersen}\ \emph
  {et~al.}(2008{\natexlab{b}})\citenamefont {Pedersen}, \citenamefont {Flindt},
  \citenamefont {Pedersen}, \citenamefont {Jauho}, \citenamefont {Mortensen},\
  and\ \citenamefont {Pedersen}}]{pedersen2008optical}%
  \BibitemOpen
  \bibfield  {author} {\bibinfo {author} {\bibfnamefont {T.~G.}\ \bibnamefont
  {Pedersen}}, \bibinfo {author} {\bibfnamefont {C.}~\bibnamefont {Flindt}},
  \bibinfo {author} {\bibfnamefont {J.}~\bibnamefont {Pedersen}}, \bibinfo
  {author} {\bibfnamefont {A.-P.}\ \bibnamefont {Jauho}}, \bibinfo {author}
  {\bibfnamefont {N.~A.}\ \bibnamefont {Mortensen}},\ and\ \bibinfo {author}
  {\bibfnamefont {K.}~\bibnamefont {Pedersen}},\ }\bibfield  {title} {\bibinfo
  {title} {Optical properties of graphene antidot lattices},\ }\href@noop {}
  {\bibfield  {journal} {\bibinfo  {journal} {Phys. Rev. B}\ }\textbf {\bibinfo
  {volume} {77}},\ \bibinfo {pages} {245431} (\bibinfo {year}
  {2008}{\natexlab{b}})}\BibitemShut {NoStop}%
\bibitem [{\citenamefont {Pohle}\ \emph {et~al.}(2018)\citenamefont {Pohle},
  \citenamefont {Kavousanaki}, \citenamefont {Dani},\ and\ \citenamefont
  {Shannon}}]{pohle2018symmetry}%
  \BibitemOpen
  \bibfield  {author} {\bibinfo {author} {\bibfnamefont {R.}~\bibnamefont
  {Pohle}}, \bibinfo {author} {\bibfnamefont {E.~G.}\ \bibnamefont
  {Kavousanaki}}, \bibinfo {author} {\bibfnamefont {K.~M.}\ \bibnamefont
  {Dani}},\ and\ \bibinfo {author} {\bibfnamefont {N.}~\bibnamefont
  {Shannon}},\ }\bibfield  {title} {\bibinfo {title} {Symmetry and optical
  selection rules in graphene quantum dots},\ }\href@noop {} {\bibfield
  {journal} {\bibinfo  {journal} {Phys. Rev. B}\ }\textbf {\bibinfo {volume}
  {97}},\ \bibinfo {pages} {115404} (\bibinfo {year} {2018})}\BibitemShut
  {NoStop}%
\bibitem [{\citenamefont {Wang}\ \emph {et~al.}(2022)\citenamefont {Wang},
  \citenamefont {Yu}, \citenamefont {R{\"o}sner}, \citenamefont {Katsnelson},
  \citenamefont {Lin},\ and\ \citenamefont {Yuan}}]{wang2022polarization}%
  \BibitemOpen
  \bibfield  {author} {\bibinfo {author} {\bibfnamefont {Y.}~\bibnamefont
  {Wang}}, \bibinfo {author} {\bibfnamefont {G.}~\bibnamefont {Yu}}, \bibinfo
  {author} {\bibfnamefont {M.}~\bibnamefont {R{\"o}sner}}, \bibinfo {author}
  {\bibfnamefont {M.~I.}\ \bibnamefont {Katsnelson}}, \bibinfo {author}
  {\bibfnamefont {H.-Q.}\ \bibnamefont {Lin}},\ and\ \bibinfo {author}
  {\bibfnamefont {S.}~\bibnamefont {Yuan}},\ }\bibfield  {title} {\bibinfo
  {title} {Polarization-dependent selection rules and optical spectrum atlas of
  twisted bilayer graphene quantum dots},\ }\href@noop {} {\bibfield  {journal}
  {\bibinfo  {journal} {Phys. Rev. X}\ }\textbf {\bibinfo {volume} {12}},\
  \bibinfo {pages} {021055} (\bibinfo {year} {2022})}\BibitemShut {NoStop}%
\bibitem [{\citenamefont {Neufeld}\ \emph {et~al.}(2019)\citenamefont
  {Neufeld}, \citenamefont {Podolsky},\ and\ \citenamefont
  {Cohen}}]{neufeld2019floquet}%
  \BibitemOpen
  \bibfield  {author} {\bibinfo {author} {\bibfnamefont {O.}~\bibnamefont
  {Neufeld}}, \bibinfo {author} {\bibfnamefont {D.}~\bibnamefont {Podolsky}},\
  and\ \bibinfo {author} {\bibfnamefont {O.}~\bibnamefont {Cohen}},\ }\bibfield
   {title} {\bibinfo {title} {Floquet group theory and its application to
  selection rules in harmonic generation},\ }\href@noop {} {\bibfield
  {journal} {\bibinfo  {journal} {Nat. Commun.}\ }\textbf {\bibinfo {volume}
  {10}},\ \bibinfo {pages} {405} (\bibinfo {year} {2019})}\BibitemShut
  {NoStop}%
\bibitem [{\citenamefont {Engelhardt}\ and\ \citenamefont
  {Cao}(2021)}]{engelhardt2021dynamical}%
  \BibitemOpen
  \bibfield  {author} {\bibinfo {author} {\bibfnamefont {G.}~\bibnamefont
  {Engelhardt}}\ and\ \bibinfo {author} {\bibfnamefont {J.}~\bibnamefont
  {Cao}},\ }\bibfield  {title} {\bibinfo {title} {Dynamical symmetries and
  symmetry-protected selection rules in periodically driven quantum systems},\
  }\href@noop {} {\bibfield  {journal} {\bibinfo  {journal} {Phys. Rev. Lett.}\
  }\textbf {\bibinfo {volume} {126}},\ \bibinfo {pages} {090601} (\bibinfo
  {year} {2021})}\BibitemShut {NoStop}%
\bibitem [{\citenamefont {Wang}\ \emph {et~al.}(2021)\citenamefont {Wang},
  \citenamefont {Li},\ and\ \citenamefont {Cappellaro}}]{wang2021observation}%
  \BibitemOpen
  \bibfield  {author} {\bibinfo {author} {\bibfnamefont {G.}~\bibnamefont
  {Wang}}, \bibinfo {author} {\bibfnamefont {C.}~\bibnamefont {Li}},\ and\
  \bibinfo {author} {\bibfnamefont {P.}~\bibnamefont {Cappellaro}},\ }\bibfield
   {title} {\bibinfo {title} {Observation of symmetry-protected selection rules
  in periodically driven quantum systems},\ }\href@noop {} {\bibfield
  {journal} {\bibinfo  {journal} {Phys. Rev. Lett.}\ }\textbf {\bibinfo
  {volume} {127}},\ \bibinfo {pages} {140604} (\bibinfo {year}
  {2021})}\BibitemShut {NoStop}%
\bibitem [{\citenamefont {Cannon}\ \emph {et~al.}(2000)\citenamefont {Cannon},
  \citenamefont {Kusmartsev}, \citenamefont {Alekseev},\ and\ \citenamefont
  {Campbell}}]{cannon2000absolute}%
  \BibitemOpen
  \bibfield  {author} {\bibinfo {author} {\bibfnamefont {E.~H.}\ \bibnamefont
  {Cannon}}, \bibinfo {author} {\bibfnamefont {F.~V.}\ \bibnamefont
  {Kusmartsev}}, \bibinfo {author} {\bibfnamefont {K.~N.}\ \bibnamefont
  {Alekseev}},\ and\ \bibinfo {author} {\bibfnamefont {D.~K.}\ \bibnamefont
  {Campbell}},\ }\bibfield  {title} {\bibinfo {title} {Absolute negative
  conductivity and spontaneous current generation in semiconductor
  superlattices with hot electrons},\ }\href@noop {} {\bibfield  {journal}
  {\bibinfo  {journal} {Phys. Rev. Lett.}\ }\textbf {\bibinfo {volume} {85}},\
  \bibinfo {pages} {1302} (\bibinfo {year} {2000})}\BibitemShut {NoStop}%
\bibitem [{\citenamefont {Ryzhii}\ and\ \citenamefont
  {Vyurkov}(2003)}]{ryzhii2003absolute}%
  \BibitemOpen
  \bibfield  {author} {\bibinfo {author} {\bibfnamefont {V.}~\bibnamefont
  {Ryzhii}}\ and\ \bibinfo {author} {\bibfnamefont {V.}~\bibnamefont
  {Vyurkov}},\ }\bibfield  {title} {\bibinfo {title} {Absolute negative
  conductivity in two-dimensional electron systems associated with acoustic
  scattering stimulated by microwave radiation},\ }\href@noop {} {\bibfield
  {journal} {\bibinfo  {journal} {Phys. Rev. B}\ }\textbf {\bibinfo {volume}
  {68}},\ \bibinfo {pages} {165406} (\bibinfo {year} {2003})}\BibitemShut
  {NoStop}%
\bibitem [{\citenamefont {I{\~n}arrea}\ and\ \citenamefont
  {Platero}(2006)}]{inarrea2006zero}%
  \BibitemOpen
  \bibfield  {author} {\bibinfo {author} {\bibfnamefont {J.}~\bibnamefont
  {I{\~n}arrea}}\ and\ \bibinfo {author} {\bibfnamefont {G.}~\bibnamefont
  {Platero}},\ }\bibfield  {title} {\bibinfo {title} {From zero resistance
  states to absolute negative conductivity in microwave irradiated
  two-dimensional electron systems},\ }\href@noop {} {\bibfield  {journal}
  {\bibinfo  {journal} {Appl. Phys. Lett.}\ }\textbf {\bibinfo {volume} {89}},\
  \bibinfo {pages} {052109} (\bibinfo {year} {2006})}\BibitemShut {NoStop}%
\bibitem [{\citenamefont {Li}\ \emph {et~al.}(2012)\citenamefont {Li},
  \citenamefont {Luo}, \citenamefont {Hupalo}, \citenamefont {Zhang},
  \citenamefont {Tringides}, \citenamefont {Schmalian},\ and\ \citenamefont
  {Wang}}]{li2012femtosecond}%
  \BibitemOpen
  \bibfield  {author} {\bibinfo {author} {\bibfnamefont {T.}~\bibnamefont
  {Li}}, \bibinfo {author} {\bibfnamefont {L.}~\bibnamefont {Luo}}, \bibinfo
  {author} {\bibfnamefont {M.}~\bibnamefont {Hupalo}}, \bibinfo {author}
  {\bibfnamefont {J.}~\bibnamefont {Zhang}}, \bibinfo {author} {\bibfnamefont
  {M.~C.}\ \bibnamefont {Tringides}}, \bibinfo {author} {\bibfnamefont
  {J.}~\bibnamefont {Schmalian}},\ and\ \bibinfo {author} {\bibfnamefont
  {J.}~\bibnamefont {Wang}},\ }\bibfield  {title} {\bibinfo {title}
  {Femtosecond population inversion and stimulated emission of dense {D}irac
  fermions in graphene},\ }\href@noop {} {\bibfield  {journal} {\bibinfo
  {journal} {Phys. Rev. Lett.}\ }\textbf {\bibinfo {volume} {108}},\ \bibinfo
  {pages} {167401} (\bibinfo {year} {2012})}\BibitemShut {NoStop}%
\bibitem [{\citenamefont {Ryzhii}\ \emph {et~al.}(2007)\citenamefont {Ryzhii},
  \citenamefont {Ryzhii},\ and\ \citenamefont {Otsuji}}]{ryzhii2007negative}%
  \BibitemOpen
  \bibfield  {author} {\bibinfo {author} {\bibfnamefont {V.}~\bibnamefont
  {Ryzhii}}, \bibinfo {author} {\bibfnamefont {M.}~\bibnamefont {Ryzhii}},\
  and\ \bibinfo {author} {\bibfnamefont {T.}~\bibnamefont {Otsuji}},\
  }\bibfield  {title} {\bibinfo {title} {Negative dynamic conductivity of
  graphene with optical pumping},\ }\href@noop {} {\bibfield  {journal}
  {\bibinfo  {journal} {J. Appl. Phys.}\ }\textbf {\bibinfo {volume} {101}},\
  \bibinfo {pages} {083114} (\bibinfo {year} {2007})}\BibitemShut {NoStop}%
\bibitem [{\citenamefont {Bandurin}\ \emph {et~al.}(2016)\citenamefont
  {Bandurin}, \citenamefont {Torre}, \citenamefont {Kumar}, \citenamefont
  {Shalom}, \citenamefont {Tomadin}, \citenamefont {Principi}, \citenamefont
  {Auton}, \citenamefont {Khestanova}, \citenamefont {Novoselov}, \citenamefont
  {Grigorieva} \emph {et~al.}}]{bandurin2016negative}%
  \BibitemOpen
  \bibfield  {author} {\bibinfo {author} {\bibfnamefont {D.~A.}\ \bibnamefont
  {Bandurin}}, \bibinfo {author} {\bibfnamefont {I.}~\bibnamefont {Torre}},
  \bibinfo {author} {\bibfnamefont {R.~K.}\ \bibnamefont {Kumar}}, \bibinfo
  {author} {\bibfnamefont {M.~B.}\ \bibnamefont {Shalom}}, \bibinfo {author}
  {\bibfnamefont {A.}~\bibnamefont {Tomadin}}, \bibinfo {author} {\bibfnamefont
  {A.}~\bibnamefont {Principi}}, \bibinfo {author} {\bibfnamefont {G.~H.}\
  \bibnamefont {Auton}}, \bibinfo {author} {\bibfnamefont {E.}~\bibnamefont
  {Khestanova}}, \bibinfo {author} {\bibfnamefont {K.~S.}\ \bibnamefont
  {Novoselov}}, \bibinfo {author} {\bibfnamefont {I.~V.}\ \bibnamefont
  {Grigorieva}}, \emph {et~al.},\ }\bibfield  {title} {\bibinfo {title}
  {Negative local resistance caused by viscous electron backflow in graphene},\
  }\href@noop {} {\bibfield  {journal} {\bibinfo  {journal} {Science}\ }\textbf
  {\bibinfo {volume} {351}},\ \bibinfo {pages} {1055} (\bibinfo {year}
  {2016})}\BibitemShut {NoStop}%
\bibitem [{\citenamefont {Levitov}\ and\ \citenamefont
  {Falkovich}(2016)}]{levitov2016electron}%
  \BibitemOpen
  \bibfield  {author} {\bibinfo {author} {\bibfnamefont {L.}~\bibnamefont
  {Levitov}}\ and\ \bibinfo {author} {\bibfnamefont {G.}~\bibnamefont
  {Falkovich}},\ }\bibfield  {title} {\bibinfo {title} {Electron viscosity,
  current vortices and negative nonlocal resistance in graphene},\ }\href@noop
  {} {\bibfield  {journal} {\bibinfo  {journal} {Nat. Phys.}\ }\textbf
  {\bibinfo {volume} {12}},\ \bibinfo {pages} {672} (\bibinfo {year}
  {2016})}\BibitemShut {NoStop}%
\bibitem [{\citenamefont {Decroly}\ \emph {et~al.}(1973)\citenamefont
  {Decroly}, \citenamefont {Laurent}, \citenamefont {Lienard}, \citenamefont
  {Marechal},\ and\ \citenamefont {Vorobeitchik}}]{decroly1973parametric}%
  \BibitemOpen
  \bibfield  {author} {\bibinfo {author} {\bibfnamefont {J.~C.}\ \bibnamefont
  {Decroly}}, \bibinfo {author} {\bibfnamefont {L.}~\bibnamefont {Laurent}},
  \bibinfo {author} {\bibfnamefont {J.~C.}\ \bibnamefont {Lienard}}, \bibinfo
  {author} {\bibfnamefont {G.}~\bibnamefont {Marechal}},\ and\ \bibinfo
  {author} {\bibfnamefont {J.}~\bibnamefont {Vorobeitchik}},\ }\href@noop {}
  {\emph {\bibinfo {title} {Parametric Amplifiers}}}\ (\bibinfo  {publisher}
  {Springer},\ \bibinfo {year} {1973})\BibitemShut {NoStop}%
\bibitem [{\citenamefont {Smith}\ \emph {et~al.}(2004)\citenamefont {Smith},
  \citenamefont {Pendry},\ and\ \citenamefont
  {Wiltshire}}]{smith2004metamaterials}%
  \BibitemOpen
  \bibfield  {author} {\bibinfo {author} {\bibfnamefont {D.~R.}\ \bibnamefont
  {Smith}}, \bibinfo {author} {\bibfnamefont {J.~B.}\ \bibnamefont {Pendry}},\
  and\ \bibinfo {author} {\bibfnamefont {M.~C.~K.}\ \bibnamefont {Wiltshire}},\
  }\bibfield  {title} {\bibinfo {title} {Metamaterials and negative refractive
  index},\ }\href@noop {} {\bibfield  {journal} {\bibinfo  {journal} {Science}\
  }\textbf {\bibinfo {volume} {305}},\ \bibinfo {pages} {788} (\bibinfo {year}
  {2004})}\BibitemShut {NoStop}%
\bibitem [{\citenamefont {Padilla}\ \emph {et~al.}(2006)\citenamefont
  {Padilla}, \citenamefont {Basov},\ and\ \citenamefont
  {Smith}}]{padilla2006negative}%
  \BibitemOpen
  \bibfield  {author} {\bibinfo {author} {\bibfnamefont {W.~J.}\ \bibnamefont
  {Padilla}}, \bibinfo {author} {\bibfnamefont {D.~N.}\ \bibnamefont {Basov}},\
  and\ \bibinfo {author} {\bibfnamefont {D.~R.}\ \bibnamefont {Smith}},\
  }\bibfield  {title} {\bibinfo {title} {Negative refractive index
  metamaterials},\ }\href@noop {} {\bibfield  {journal} {\bibinfo  {journal}
  {Mater. Today}\ }\textbf {\bibinfo {volume} {9}},\ \bibinfo {pages} {28}
  (\bibinfo {year} {2006})}\BibitemShut {NoStop}%
\bibitem [{\citenamefont {Shalaev}(2007)}]{shalaev2007optical}%
  \BibitemOpen
  \bibfield  {author} {\bibinfo {author} {\bibfnamefont {V.~M.}\ \bibnamefont
  {Shalaev}},\ }\bibfield  {title} {\bibinfo {title} {Optical negative-index
  metamaterials},\ }\href@noop {} {\bibfield  {journal} {\bibinfo  {journal}
  {Nat. Photonics}\ }\textbf {\bibinfo {volume} {1}},\ \bibinfo {pages} {41}
  (\bibinfo {year} {2007})}\BibitemShut {NoStop}%
\bibitem [{\citenamefont {Kadic}\ \emph {et~al.}(2019)\citenamefont {Kadic},
  \citenamefont {Milton}, \citenamefont {van Hecke},\ and\ \citenamefont
  {Wegener}}]{kadic20193d}%
  \BibitemOpen
  \bibfield  {author} {\bibinfo {author} {\bibfnamefont {M.}~\bibnamefont
  {Kadic}}, \bibinfo {author} {\bibfnamefont {G.~W.}\ \bibnamefont {Milton}},
  \bibinfo {author} {\bibfnamefont {M.}~\bibnamefont {van Hecke}},\ and\
  \bibinfo {author} {\bibfnamefont {M.}~\bibnamefont {Wegener}},\ }\bibfield
  {title} {\bibinfo {title} {3{D} metamaterials},\ }\href@noop {} {\bibfield
  {journal} {\bibinfo  {journal} {Nat. Rev. Phys.}\ }\textbf {\bibinfo {volume}
  {1}},\ \bibinfo {pages} {198} (\bibinfo {year} {2019})}\BibitemShut {NoStop}%
\bibitem [{\citenamefont {Bukov}\ and\ \citenamefont
  {Polkovnikov}(2014)}]{bukov2014}%
  \BibitemOpen
  \bibfield  {author} {\bibinfo {author} {\bibfnamefont {M.}~\bibnamefont
  {Bukov}}\ and\ \bibinfo {author} {\bibfnamefont {A.}~\bibnamefont
  {Polkovnikov}},\ }\bibfield  {title} {\bibinfo {title} {Stroboscopic versus
  nonstroboscopic dynamics in the {F}loquet realization of the
  {H}arper-{H}ofstadter {H}amiltonian},\ }\href
  {https://doi.org/10.1103/PhysRevA.90.043613} {\bibfield  {journal} {\bibinfo
  {journal} {Phys. Rev. A}\ }\textbf {\bibinfo {volume} {90}},\ \bibinfo
  {pages} {043613} (\bibinfo {year} {2014})}\BibitemShut {NoStop}%
\bibitem [{\citenamefont {Goldman}\ and\ \citenamefont
  {Dalibard}(2014)}]{goldman2014}%
  \BibitemOpen
  \bibfield  {author} {\bibinfo {author} {\bibfnamefont {N.}~\bibnamefont
  {Goldman}}\ and\ \bibinfo {author} {\bibfnamefont {J.}~\bibnamefont
  {Dalibard}},\ }\bibfield  {title} {\bibinfo {title} {Periodically driven
  quantum systems: {E}ffective {H}amiltonians and engineered gauge fields},\
  }\href {https://doi.org/10.1103/PhysRevX.4.031027} {\bibfield  {journal}
  {\bibinfo  {journal} {Phys. Rev. X}\ }\textbf {\bibinfo {volume} {4}},\
  \bibinfo {pages} {031027} (\bibinfo {year} {2014})}\BibitemShut {NoStop}%
\bibitem [{\citenamefont {Castro}\ \emph {et~al.}(2022)\citenamefont {Castro},
  \citenamefont {De~Giovannini}, \citenamefont {Sato}, \citenamefont
  {H\"ubener},\ and\ \citenamefont {Rubio}}]{rubio2022OCT}%
  \BibitemOpen
  \bibfield  {author} {\bibinfo {author} {\bibfnamefont {A.}~\bibnamefont
  {Castro}}, \bibinfo {author} {\bibfnamefont {U.}~\bibnamefont
  {De~Giovannini}}, \bibinfo {author} {\bibfnamefont {S.~A.}\ \bibnamefont
  {Sato}}, \bibinfo {author} {\bibfnamefont {H.}~\bibnamefont {H\"ubener}},\
  and\ \bibinfo {author} {\bibfnamefont {A.}~\bibnamefont {Rubio}},\ }\bibfield
   {title} {\bibinfo {title} {Floquet engineering the band structure of
  materials with optimal control theory},\ }\href
  {https://doi.org/10.1103/PhysRevResearch.4.033213} {\bibfield  {journal}
  {\bibinfo  {journal} {Phys. Rev. Research}\ }\textbf {\bibinfo {volume}
  {4}},\ \bibinfo {pages} {033213} (\bibinfo {year} {2022})}\BibitemShut
  {NoStop}%
\end{thebibliography}%


\end{document}